\documentclass[iop]{emulateapj}

\usepackage{color}
\usepackage[outdir=./]{epstopdf}
\usepackage[table]{xcolor}

\shorttitle{Unveiling the buried nucleus in NGC 1448 with \emph{N\MakeLowercase{u}STAR}}
\shortauthors{A. Annuar et al.}

\begin{document}


\title{A New Compton-thick AGN in our Cosmic Backyard: Unveiling the Buried Nucleus in NGC 1448 with \emph{N\MakeLowercase{u}STAR}}

\author{A. Annuar\altaffilmark{1}, D. M. Alexander\altaffilmark{1}, P. Gandhi\altaffilmark{2}, G. B. Lansbury\altaffilmark{1}, D. Asmus\altaffilmark{3}, D. R. Ballantyne\altaffilmark{4}, F. E. Bauer\altaffilmark{5,6,7},  S. E. Boggs\altaffilmark{8}, P. G. Boorman\altaffilmark{2}, W. N. Brandt\altaffilmark{9,10,11}, M. Brightman\altaffilmark{12}, F. E. Christensen\altaffilmark{13}, W. W. Craig\altaffilmark{8,14}, D. Farrah\altaffilmark{15}, A. D. Goulding\altaffilmark{16}, C. J. Hailey\altaffilmark{17}, F. A. Harrison\altaffilmark{12}, M. J. Koss\altaffilmark{18}, S. M. LaMassa\altaffilmark{19}, S. S. Murray\altaffilmark{$\dagger$,20,21}, C. Ricci\altaffilmark{5,22}, D. J. Rosario\altaffilmark{1}, F. Stanley\altaffilmark{1}, D. Stern\altaffilmark{23} and W. Zhang\altaffilmark{19}.}

\affil{$^{1}$Centre for Extragalactic Astronomy, Department of Physics, Durham University, South Road, Durham, DH1 3LE, UK}
\affil{$^{2}$Department of Physics $\&$ Astronomy, Faculty of Physical Sciences and Engineering, University of Southampton, Southampton, SO17 1BJ, UK}
\affil{$^{3}$European Southern Observatory, Alonso de Cordova, Vitacura, Casilla 19001, Santiago, Chile}
\affil{$^{4}$Center for Relativistic Astrophysics, School of Physics, Georgia Institute of Technology, Atlanta, GA 30332, USA}
\affil{$^{5}$Instituto de Astrof{\'{\i}}sica and Centro de Astroingenier{\'{\i}}a, Facultad de F{\'{i}}sica, Pontificia Universidad Cat{\'{o}}lica de Chile, Casilla 306, Santiago 22, Chile} 
\affil{$^{6}$Millennium Institute of Astrophysics (MAS), Nuncio Monse{\~{n}}or S{\'{o}}tero Sanz 100, Providencia, Santiago, Chile} 
\affil{$^{7}$Space Science Institute, 4750 Walnut Street, Suite 205, Boulder, Colorado 80301} 
\affil{$^{8}$Space Sciences Laboratory, University of California, Berkeley CA 94720, USA}
\affil{$^{9}$Department of Astronomy and Astrophysics, The Pennsylvania State University, 525 Davey Lab, University Park, PA 16802, USA}
\affil{$^{10}$Institute for Gravitation and the Cosmos, The Pennsylvania State University, University Park, PA 16802, USA}
\affil{$^{11}$Department of Physics, The Pennsylvania State University, 525 Davey Lab, University Park, PA 16802, USA}
\affil{$^{12}$Cahill Center for Astronomy and Astrophysics, California Institute of Technology, Pasadena, CA 91125, USA}
\affil{$^{13}$DTU Space, National Space Institute, Technical University of Denmark, Elektrovej 327, DK-2800 Lyngby, Denmark}
\affil{$^{14}$Lawrence Livermore National Laboratory, Livermore, CA 94550, USA}
\affil{$^{15}$Department of Physics, Virginia Tech, Blacksburg, VA 24061, USA}
\affil{$^{16}$Department of Astrophysical Sciences, Princeton University, Princeton, NJ 08544, USA}
\affil{$^{17}$Columbia Astrophysics Laboratory, Columbia University, New York, NY 10027, USA}
\affil{$^{18}$Institute for Astronomy, Department of Physics, ETH Zurich, Wolfgang-Pauli-Strasse 27, CH-8093 Zurich, Switzerland}
\affil{$^{19}$NASA Goddard Space Flight Center, Greenbelt, Maryland 20771, USA}
\affil{$^{20}$Harvard-Smithsonian Center for Astrophysics, 60 Garden Street, Cambridge, MA 02138, USA}
\affil{$^{21}$Department of Physics and Astronomy, Johns Hopkins University, 3400 North Charles Street, Baltimore, MD 21218, USA}
\affil{$^{22}$Kavli Institute for Astronomy and Astrophysics, Peking University, Beijing 100871, China}
\affil{$^{23}$Jet Propulsion Laboratory, California Institute of Technology, Pasadena, CA 91109, USA}

%\affil{$^{18}$Instituto de Astrofísica, Facultad de Física, Pontificia Universidad Católica de Chile, Casilla 306, Santiago 22, Chile}
%%%%%%%%%%%%%%%%%%%%%%%%%%%%%%%%%%%%%%%%%%%%%%%%%%%%%%%%%%%%%%%%%%%%%%%%%%%%%%%%%%%%%%%%%%%%%%%%%%%%%%%%%%%

\begin{abstract}
NGC 1448 is one of the nearest luminous galaxies ($L_{8-1000\mu m} >$ 10$^{9} L_{\odot}$) to ours ($z$ $=$ 0.00390), and yet the active galactic nucleus (AGN) it hosts was only recently discovered, in 2009. In this paper, we present an analysis of the nuclear source across three wavebands: mid-infrared (MIR) continuum, optical, and X-rays. We observed the source with the {\em{Nuclear Spectroscopic Telescope Array}} (\textsl{NuSTAR}), and combined this data with archival \textsl{Chandra} data to perform broadband X-ray spectral fitting ($\approx$0.5--40 keV) of the AGN for the first time. Our X-ray spectral analysis reveals that the AGN is buried under a Compton-thick (CT) column of obscuring gas along our line-of-sight, with a column density of \emph{N}$_{\rm H}$(\rm los) $\gtrsim$ 2.5 $\times$ 10$^{24}$ cm$^{-2}$. The best-fitting torus models measured an intrinsic 2--10 keV luminosity of \emph{L}$_{2-10\rm{,int}}$ $=$ (3.5--7.6) $\times$ 10$^{40}$ erg s$^{-1}$, making NGC 1448 one of the lowest luminosity CTAGNs known. In addition to the \textsl{NuSTAR} observation, we also performed optical spectroscopy for the nucleus in this edge-on galaxy using the European Southern Observatory New Technology Telescope. We re-classify the optical nuclear spectrum as a Seyfert on the basis of the Baldwin-Philips-Terlevich diagnostic diagrams, thus identifying the AGN at optical wavelengths for the first time. We also present high spatial resolution MIR observations of NGC 1448 with Gemini/T-ReCS, in which a compact nucleus is clearly detected. The absorption-corrected 2--10 keV luminosity measured from our X-ray spectral analysis agrees with that predicted from the optical [O{\sc{iii}}]$\lambda$5007\AA \ emission line and the MIR 12$\micron$ continuum, further supporting the CT nature of the AGN. 
\end{abstract}


%%%%%%%%%%%%%%%%%%%%%%%%%%%%%%%%%%%%%%%%%%%%%%%%%%%%%%%%%%%%%%%%%%%%%%%%%%%%%%%%%%%%%%%%%%%%%%%%%%%%%%%%%%%

\section{Introduction}

{\let\thefootnote\relax\footnote{$\dagger$ Deceased.}}At a distance of 11.5 Mpc ($z$ $=$ 0.00390),$^{1}$\let\thefootnote\relax\footnote{$^{1}$ The quoted distance is the metric/proper distance calculated based upon the \citet{Mould00} cosmic attractor model, using $H_{\rm 0} =$ 75 km s$^{-1}$ Mpc$^{-1}$ and adopting flat cosmology ($\Omega_{M} =$ 0.3, $\Omega_{\Lambda} =$ 0.7, $q_{\rm 0} =$ 0.3) \citep{Sanders03}.} NGC 1448 is one of the nearest bolometrically luminous galaxies ($L_{8-1000\mu m} >$ 10$^{9} L_{\odot}$) to our own. Yet, the presence of an active galactic nucleus (AGN) at the center of the galaxy was only discovered less than a decade ago by \citet{Goulding09} through the detection of the high-ionization [Ne{\sc{v}}]$\lambda$14.32$\mu$m emission line, as part of a sample of luminous galaxies observed by \emph{Spitzer}. Based on the [Ne{\sc{v}}]/$f_{8-1000\mu m}$ and [Ne{\sc{v}}]/[Ne{\sc{ii}}]$\lambda$12.82$\mu$m luminosity ratios, they found that the AGN contributes a significant fraction ($>$25$\%$) of the infrared (IR) emission. So far, the focus of the majority of studies of NGC 1448 has been on a number of supernovae occuring in the galaxy (e.g.; \citealp{Wang03}; \citealp{Sollerman05}; \citealp{Monard14}).

Based on its total $K$-band magnitude from the 2MASS Large Galaxy Atlas, $K_{\rm Tot} =$ 7.66 \citep{Jarrett03}, the total mass of NGC 1448 is log ($M_{\rm gal}$/$M_{\odot}$) $=$ 10.3 (using the stellar mass-to-light ratio versus $B - V$ relation of \citealp{Bell03}.) The star formation rate (SFR) of the galaxy estimated from its far-IR luminosity measured by the \textsl{Infrared Astronomical Satellite} (\textsl{IRAS}), log $L_{\rm fir}=$ 9.70 $L_{\odot}$ \citep{Sanders03}, is SFR $\sim$ 1 $M_{\odot}$/yr \citep{Kennicutt98}. This is consistent with the SFR expected at this redshift, given the mass of the galaxy \citep{Dave08}. The nucleus is classified as an H{\sc{ii}} region in the optical by \citet{Veron86} using the H$\alpha$ to [N{\sc{ii}}]$\lambda$6583\AA \ line ratio (H$\alpha$/[N{\sc{ii}}] $>$ 1.7), without any clear evidence for an AGN. The [O{\sc{iii}}]$\lambda$5007\AA \ and H$\beta$ emission lines were not detected in their observation, which could be attributed to obscuration by the highly inclined Scd host galaxy (\emph{i} $\approx$ 86$^{\circ}$ relative to the plane of the sky).$^{2}$\let\thefootnote\relax\footnote{$^{2}$ We obtained the host galaxy inclination from the HyperLeda website (http://leda.univ-lyon1.fr/).} Based on the [O{\sc{iv}}]$\lambda$25.89$\mu$m to [O{\sc{iii}}]$\lambda$5007\AA \ lower limit emission line ratio, \citet{Goulding09} found an extinction of $A_{V} >$ 5 mag within the galaxy, suggesting high obscuration towards the AGN. 

In recent years, many studies have been done to find the most heavily obscured AGNs, particularly Compton-thick (CT) AGNs, in the local universe. CTAGNs are AGNs that are obscured along our line-of-sight by gas with column density of $N_{\rm{H}}$ $\geq$ 1/$\sigma_{T}$ $=$ 1.5 $\times$ 10$^{24}$ cm$^{-2}$ (where $\sigma_{T}$ is the Thomson scattering cross-section). The obscuring gas is predominantly attributed to the AGN circumnuclear torus posited by the AGN unification model (\citealp{Antonucci93}; \citealp{Urry95}), but can also be contributed by larger-scale molecular clouds and dust lanes (e.g., \citealp{Elvis12}; \citealp{Prieto14}). The high column of gas severely absorbs the direct X-ray emission from the AGN, even at E $\gtrsim$ 10 keV in cases where the absorption is extreme ($N_{\rm{H}}$ $>$ 10$^{25}$ cm$^{-2}$; see Figure 1 of \citealp{Gilli07}). This is what makes it very challenging to identify CTAGNs. Indeed, a hard X-ray study ($E >$ 10 keV) by \citet{Ricci16} using a large AGN sample from the \textsl{Swift} Burst Alert Telescope (BAT) 70-month survey catalog, found an observed CTAGN fraction of $\sim$8$\%$ at $z \simeq$ 0.055. This is significantly lower than that expected from the synthesis of the cosmic X-ray background spectrum (10$\%$--25$\%$; e.g., \citealp{Gilli07}; \citealp{Treister09}; \citealp{DraperBallantyne10}; \citealp{Akylas12}; \citealp{Ueda14}), and what is predicted from multiwavelength studies of nearby AGN ($\sim$30$\%$; e.g., \citealp{Risaliti99}; \citealp{Goulding11}). This suggests that we are still missing a significant number of CTAGN, even in the local universe. Forming a complete census of their population is important in understanding the cosmic X-ray background spectrum and the growth of supermassive black holes. 

\begin{figure*}
\epsscale{0.7}
\plotone{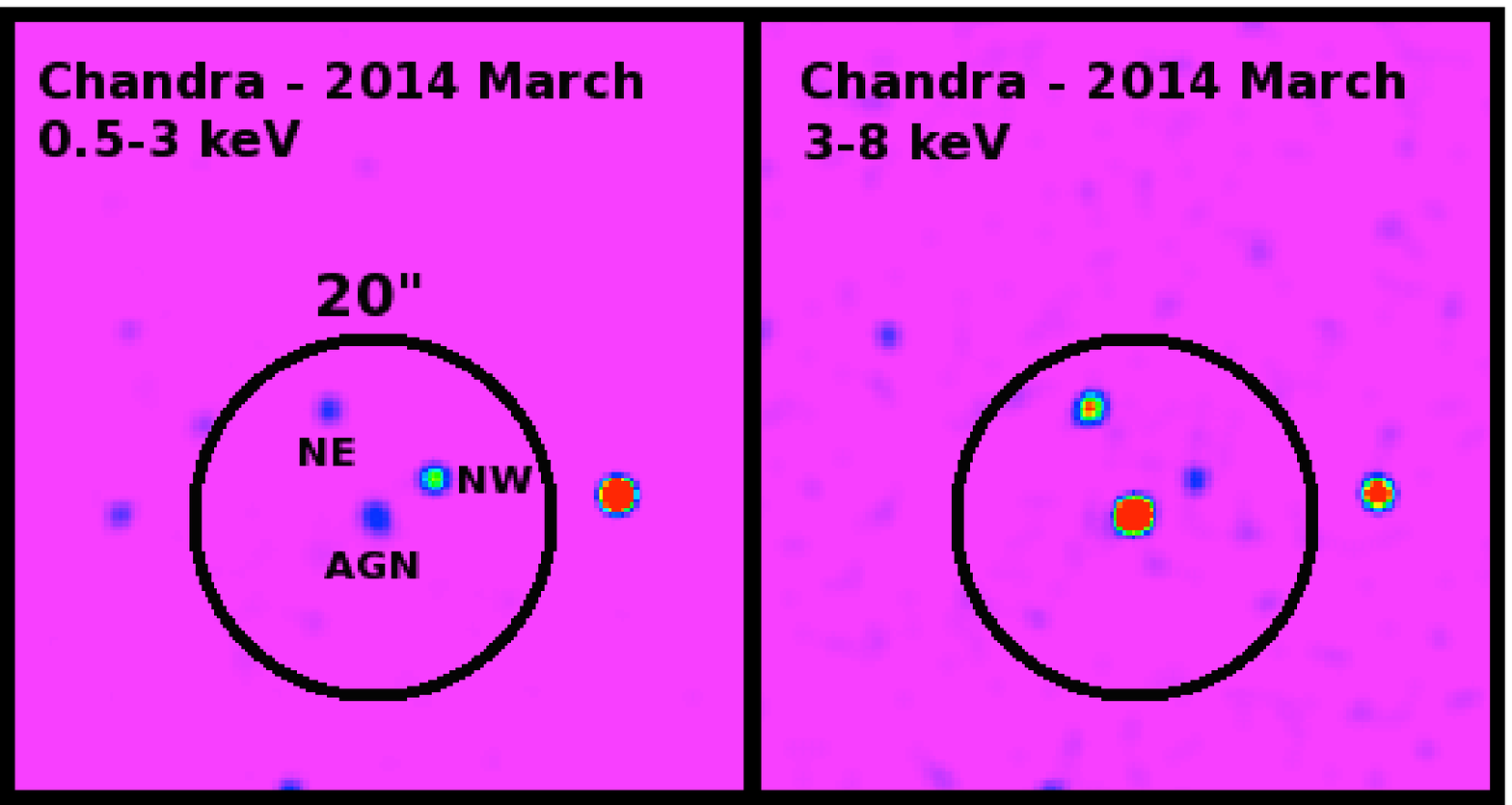}
\epsscale{0.02125}
\plotone{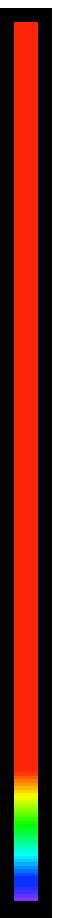}
\epsscale{1.0}
\plotone{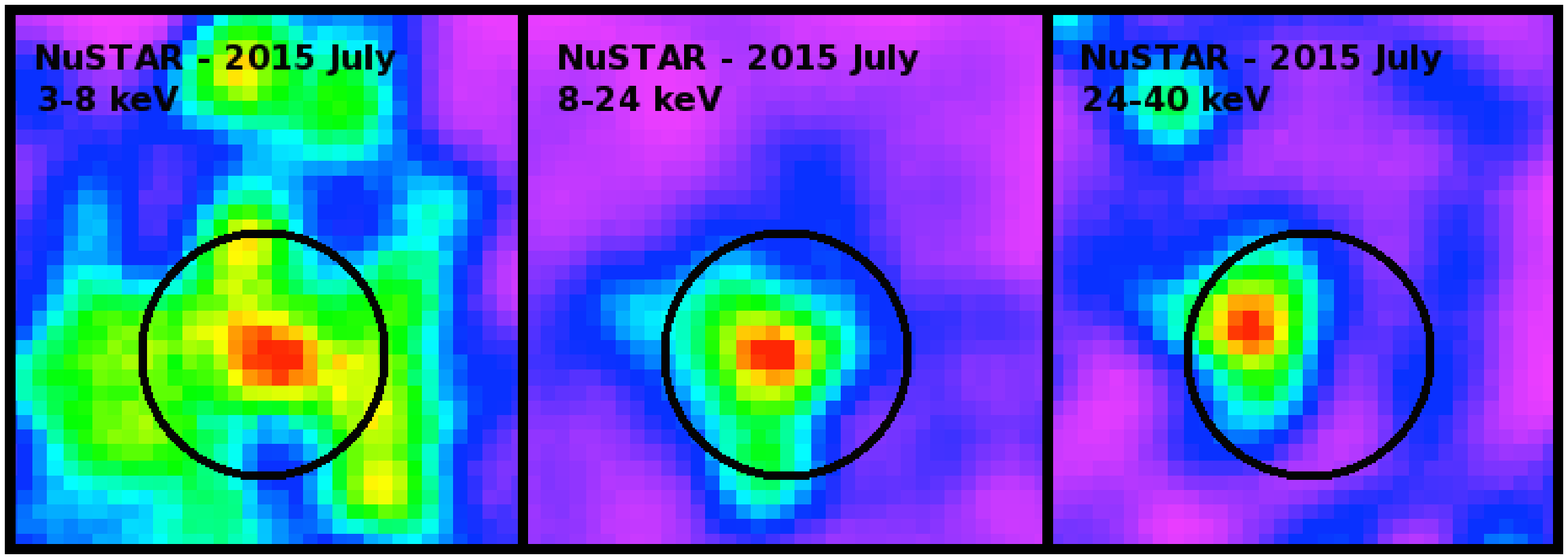}
\epsscale{0.0241}
\plotone{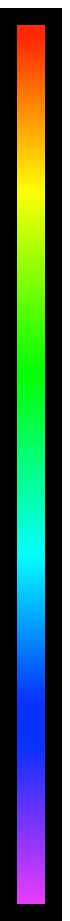}
\caption{\textsl{Chandra} images of NGC 1448 in the 0.5--3 keV and 3--8 keV bands (top panel) and \textsl{NuSTAR} FPMA+B images in the 3--8 keV, 8--24 keV and 24--40 keV bands (bottom panel). The color scale for \textsl{Chandra} and \textsl{NuSTAR} images are shown with magenta and red representing the lowest and highest counts in each image, respectively. The black circle marks a 20$\arcsec$-radius region centered on the \textsl{Chandra} position of the AGN that was used to extract the X-ray spectra. Images are smoothed with a Gaussian function of radius 5 pixels, corresponding to 2.47$\arcsec$ and 12.3$\arcsec$ for \textsl{Chandra} and \textsl{NuSTAR}, respectively. North is up and east is to the left in all images. The two off-nuclear X-ray sources detected within the extraction region in the \textsl{Chandra} image are at the north-east and north-west of the AGN, labeled as NE and NW, respectively.}
\end{figure*}

In this paper, we present \textsl{NuSTAR} and \textsl{Chandra} observations of NGC 1448 in which the AGN is detected in X-rays for the first time. We find that our direct view towards the AGN is hindered by a Compton-thick column of obscuring gas. We also report the results of new high angular resolution mid-infrared (MIR) and optical observations of the source by Gemini/T-ReCS and NTT/EFOSC2, respectively, which also reveal the presence of a buried AGN. We organized the paper as follows: In Section 2, we describe details of the multiwavelength observations and data reduction procedures for NGC 1448. We present the X-ray spectral modeling and results in Section 3, followed by the data analysis and results of the optical and MIR observations in Section 4. Finally, we discuss and summarize the overall results in Section 5 and Section 6, respectively.

\section{Observations}

In this section, we describe the new \textsl{NuSTAR} (Sections 2.1) and archival \textsl{Chandra} (Sections 2.2) observations of NGC 1448. The \textsl{NuSTAR} data are essential for tracing the intrinsic emission from the buried AGN at high X-ray energies ($E \gtrsim$ 10 keV). The \textsl{Chandra} data were used to aid our X-ray spectral analysis of the AGN at lower energies ($E \lesssim$ 3 keV) where \textsl{NuSTAR} is not sensitive, and to reliably account for the emission from contaminating off-nuclear X-ray sources in the \textsl{NuSTAR} spectrum. Combining both \textsl{NuSTAR} and \textsl{Chandra} data together allows us to analyze the X-ray spectrum of the AGN in NGC 1448 over a broadband range of energy for the first time. We also describe new optical data (spectroscopy and imaging) obtained with the ESO NTT (Section 2.3), and the new high angular resolution MIR observations with the Gemini-South telescope (Section 2.4), to provide a multiwavelength view of the AGN.

In addition to these data, we note that NGC 1448 has also been observed by \textsl{Suzaku} in X-rays for an exposure time of $\approx$ 53 ks in 2009 (2009-02-17; PI D. M. Alexander; ObsID 703062010). However, it is only significantly detected in the X-ray Imaging Spectrometer (XIS) instrument up to $\sim$10 keV. Analysis of the deepest \textsl{Swift}-BAT 104-month maps using custom detection techniques to look for faint sources \citep{Koss13}, show no significant excess (signal-to-noise ratio, SNR $= -$0.1) in the area near NGC 1448. The source was only detected in 1 out of 5 \textsl{Swift} X-ray Telescope (XRT) observations with $\sim$12 counts in $\sim$10 ks exposure time (2009-11-28; ObsID 00031031001). We do not include these data in our X-ray analysis as they do not provide additional constraints beyond those already achieved with our \textsl{NuSTAR} and \textsl{Chandra} data.

\subsection{\textsl{NuSTAR}}

\begin{figure*}
\epsscale{0.95}
\plotone{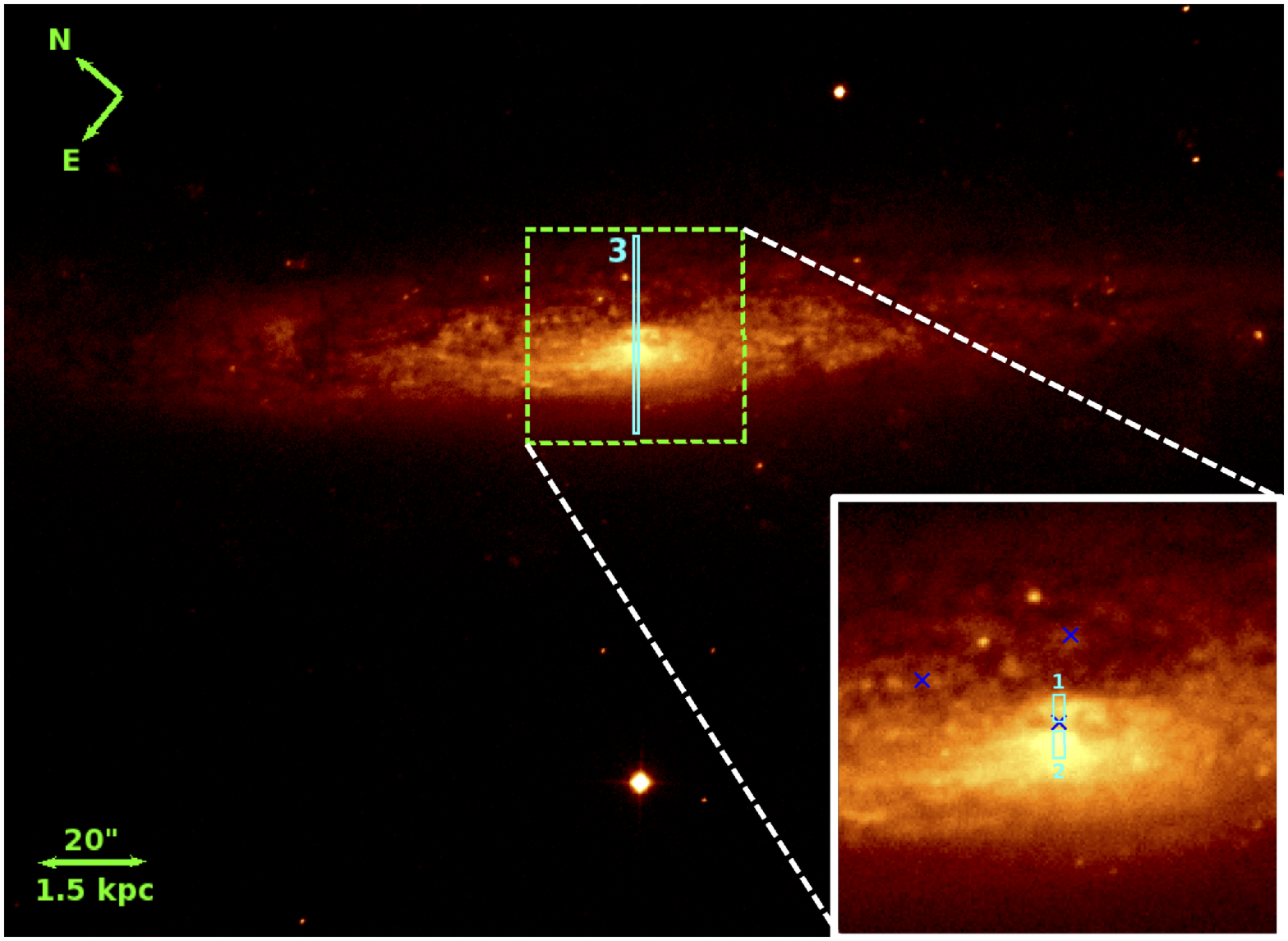}
\caption{Optical R-band image of NGC 1448 taken with the ESO NTT. The aperture used to extract the optical spectra of the AGN, optical peak and the total galaxy are plotted using cyan rectangle regions, and labeled as 1, 2 and 3, respectively. The zoom-in image of the central 20$\arcsec$ $\times$ 20$\arcsec$ of the galaxy is shown in the bottom-right panel. ``$\times$'' marks the sources detected within a 20$\arcsec$ circular radius of the AGN in the \textsl{Chandra} 2--8 keV band image.}
\end{figure*}

\begin{figure}
\epsscale{0.5}
\plotone{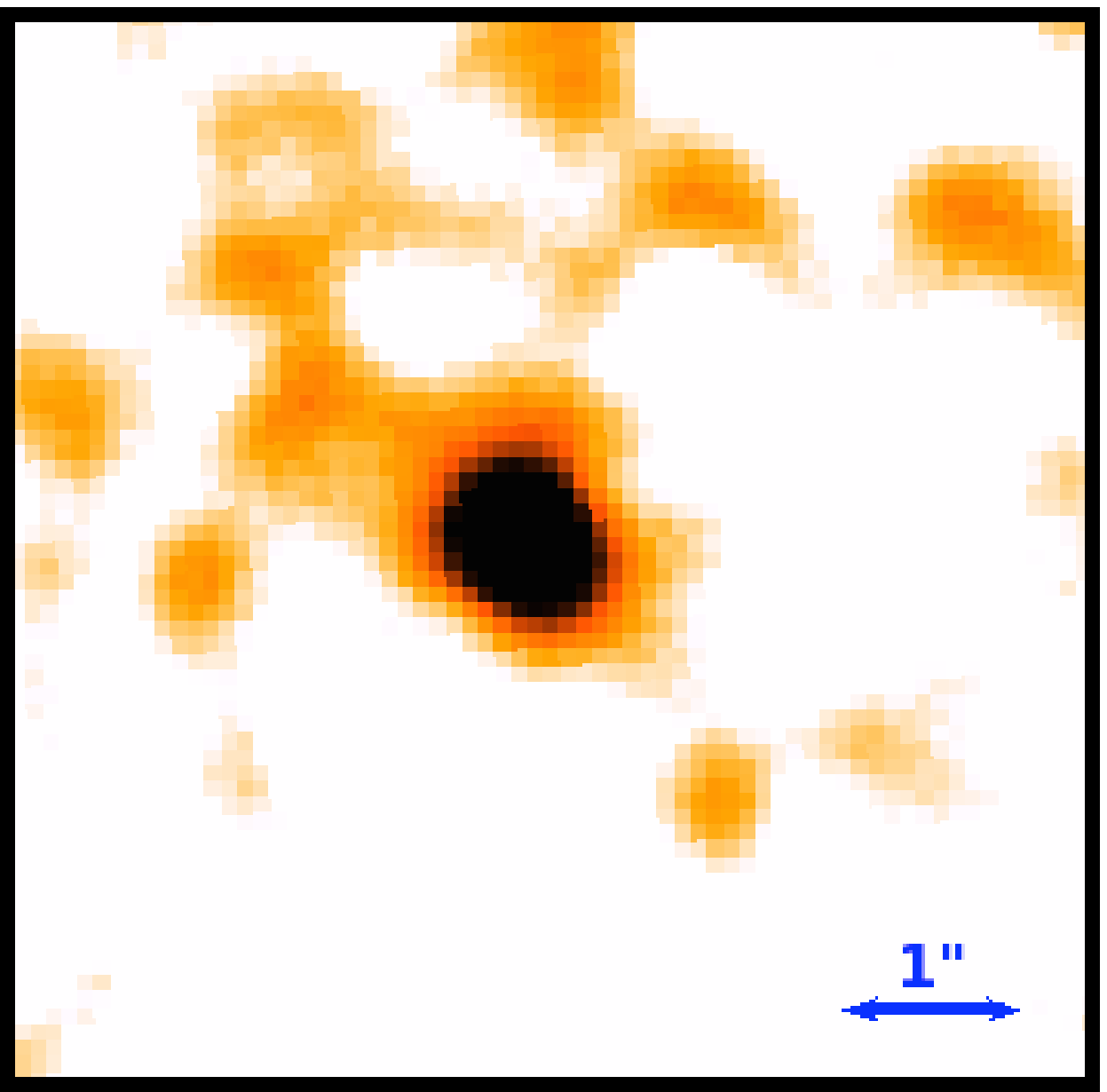}
\epsscale{0.8}
\plotone{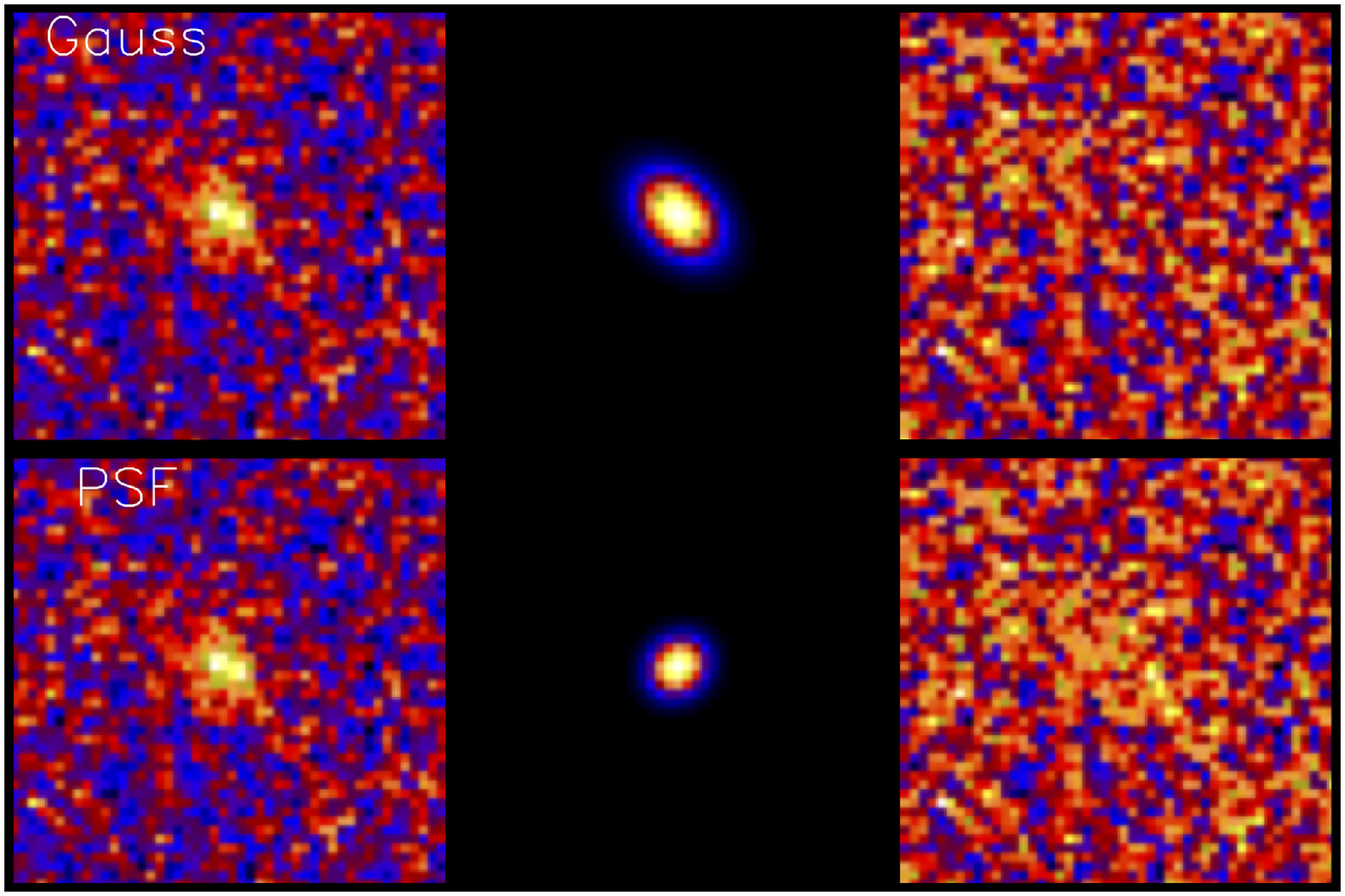}
\caption{Top: High spatial resolution MIR image of NGC 1448 taken by Gemini/T-ReCS. Bottom: Image fits performed using the {\sc{mirphot}} task in {\sc{idl}} following \citet{Asmus14}. Shown are the central 4$\farcs$5 $\times$ 4$\farcs$5 of the image. The left column is the original image, the middle column is the fit performed on the data, and the right column is the residual after subtracting the fit from the data. The top row shows the results from a Gaussian fit to the total emission, and the bottom row shows the results from fitting the standard star (as a PSF reference) to the total emission. In all images, North is up and east is to the left.}
\end{figure}

The \textsl{Nuclear Spectroscopic Telescope Array} (\textsl{NuSTAR}; \citealp{Harrison13}), launched in 2012 June, is the first orbiting X-ray observatory which focus light at high energies ($E$ $\gtrsim$ 10 keV). It consists of two co-aligned focal plane modules (FPMs), which are identical in design, and referred to as FPMA and FPMB. Each FPM covers the same 12$\arcmin$ $\times$ 12$\arcmin$ portion of the sky, and comprises four Cadmium-Zinc-Telluride detectors placed in a 2 $\times$ 2 array. \textsl{NuSTAR} operates between 3 and 79 keV, and provides 100$\times$ improvement in sensitivity as compared to previous hard X-ray orbiting observatories ($E \gtrsim$ 10 keV; e.g., \textsl{Swift}-BAT and \textsl{INTEGRAL}). In addition, it has an angular resolution of 18$\arcsec$ full width at half maximum (FWHM), with a half power diameter of 58$\arcsec$, resulting in 10$\times$ improvement over previous observatories operating at $E \gtrsim$ 10 keV. These characteristics, particularly its high energy coverage and better sensitivity, have allowed us to perform accurate broadband X-ray spectral modelings of heavily obscured AGNs in the local universe (e.g., \citealp{Balokovic14}; \citealp{Puccetti14}; \citealp{Bauer15}; \citealp{Annuar15}; \citealp{Gandhi16}).

NGC 1448 was observed by \emph{NuSTAR} in 2015 (2015-07-12; ObsID 60101101002) with an effective exposure time of 58.9 ks (60.3 ks on-source time) for each FPM. The source was observed as part of our program to form the most complete census of the CTAGN population and the $N_{\rm{H}}$ distribution of AGN in the local universe, using a volume-limited ($D <$ 15 Mpc) AGN sample from \citet{Goulding09}.$^{3}$\let\thefootnote\relax\footnote{$^{3}$ The results of the first source in the sample observed by \textsl{NuSTAR} as part of this program, NGC 5643, was reported in \citet{Annuar15}.}

We processed the {\emph{NuSTAR}} data of NGC 1448 with the \textsl{NuSTAR} Data Analysis Software ({\sc{nustardas}}) v1.4.1 within {\sc{heasoft}} v6.15.1 with CALDB v20150316. The {\sc{nupipeline}} v0.4.3 script were used to produce the calibrated and cleaned event files using standard filter flags. We extracted the spectra and response files using the {\sc{nuproducts}} v0.2.5 task.

The AGN is detected in both of the \textsl{NuSTAR} FPMs. standard filter flags. The combined images of the AGN from the two FPMs in the 3--8, 8--24, and 24--40 keV bands are shown in Figure 1. We extracted the \textsl{NuSTAR} spectrum of NGC 1448 from each FPM using a circular aperture region of 20$\arcsec$-radius (corresponding to $\sim$30$\%$ \textsl{NuSTAR} encircled energy fraction, ECF) centered on the \textsl{Chandra} position of the AGN (see Section 2.2). The aperture size was chosen to minimize contamination from off-nuclear sources observed in the \textsl{Chandra} data. The background photons were collected from an annulus region centered on the AGN with inner and outer radii of 40$\arcsec$ and 70$\arcsec$, respectively. The extracted spectrum from each FPM were then co-added using the {\sc{addascaspec}} script to increase the overall SNR of the data. We detected significant counts up to $\sim$40 keV from this combined spectrum, and measured a net count rate of 2.79 $\times$ 10$^{-3}$ counts s$^{-1}$ in the 3--40 keV band. 

We note that in the 24--40 keV band image, where we expect the AGN to completely dominate, the peak emission appears to be offset from the center of the extraction region. The offset is not observed in the 3--8 and 8--24 keV band images where {\emph{NuSTAR}} is more sensitive. Performing our spectral fits up to only 24 keV gave consistent results with that obtained by the 0.5--40 keV spectral fits. We therefore attribute this offset due to {\emph{NuSTAR}} statistical uncertainty due to lower SNR at this higher energy band. We note that centering the {\emph{NuSTAR}} extraction region according to this offset, or enlarging the extraction region to account for this apparent offset also do not significantly affect our final results on the analysis of the AGN in NGC 1448.

\subsection{\textsl{Chandra}}

NGC 1448 was observed by \textsl{Chandra} in 2014 with an exposure time of 49.4 ks (50.1 ks on-source time) using the ACIS-S detector as part of the \textsl{Chandra} HRC-GTO program (2014-03-09; PI S. Murray;  ObsID 15332). We reprocessed the data to create event files with updated calibration modifications, following standard procedures.

We determined the centroid position of the AGN in the \textsl{Chandra} hard energy band of 2--8 keV using the {\sc{wavdetect}} tool within {\sc{ciao}} with the threshold parameter set to 1 $\times$ 10$^{-7}$. We detected three sources within the central 20$\arcsec$-radius of the galaxy in this energy band (see Figure 1). The brightest source was detected at position of RA $=$ 3:44:31.83, and Dec. $=$ $-$44:38:41.22, with errors of 0$\farcs$17 and 0$\farcs$11, respectively. This is consistent with the 2MASS and Gemini/T-ReCS (see Section 2.4) positions of the nucleus within $\sim$1$\arcsec$. Therefore, we adopted this \textsl{Chandra} position as the AGN position. 

We extracted the source spectrum using the {\sc{specextract}} task in {\sc{ciao}} from a circular region of 20$\arcsec$-radius centered on the detected position of the AGN to match the \textsl{NuSTAR} extraction region. A 50$\arcsec$-radius circular aperture was used to extract the background counts from an offset, source-free region. The total net count rate within the 20$\arcsec$-radius extraction region in the 0.5--8 keV band is 5.69 $\times$ 10$^{-3}$ counts s$^{-1}$. The net count rate measured by {\sc{wavdetect}} for the AGN is 1.39 $\times$ 10$^{-3}$ counts s$^{-1}$ in the 0.5--8 keV band. The two other sources detected within the extraction region are located to the north-east (NE) and north-west (NW) of the AGN. They do not have counterparts at other wavelengths, and are likely to be X-ray binaries within NGC 1448 (see Section 3.2.1). These sources have 0.5--8 keV count rates of 6.67 $\times$ 10$^{-4}$ counts s$^{-1}$ and 8.61 $\times$ 10$^{-4}$ counts s$^{-1}$, for the NE and NW sources, respectively, as measured by {\sc{wavdetect}}.

\subsection{NTT/EFOSC2}

At optical wavelengths, we performed spectroscopy for NGC 1448 using the European Southern Observatory (ESO) New Technology Telescope (NTT) on 2015-12-07, using the Faint Object Spectrograph and Camera v.2 (EFOSC2) instrument (Program ID 096.B-0947(A); PI G. B. Lansbury). The source was observed for four 5 min exposures centered on the \textsl{Chandra} position (see Section 2.2). The slit adopted was 1$\arcsec$ (75 pc) in width, and the adopted grism yielded a spectral coverage of 4085--7520\AA, and spectral resolution of FWHM $=$ 12.6\AA. The seeing at the time of the observation was $\sim$0$\farcs$7. Standard {\sc{iraf}} routines were followed to reduce the spectra, and spectrophotometric standard star observations from the same night were used for calibration. As shown in Figure 2, we extracted spectra from three different apertures along the slit, corresponding to: (1) the AGN (using the \textsl{Chandra} position as a reference); (2) the ``optical peak"; and (3) the total galaxy. The optical peak corresponds to the position where the optical emission is the brightest in the imaging data and in the 2-dimensional spectra, and is offset from the AGN position by $\sim$3$\arcsec$ (RA $=$ 3:44:31.99 and Dec. $=$ $-$44:38:42.33). The spatial apertures adopted were 2$\farcs$4 for the AGN and the optical peak, and 36$\farcs$9 for the total galaxy (see Figure 2). The apertures for the AGN and the optical peak correspond to two spatially distinct, prominent line-emitting regions which are separated by a dust lane, from which we do not see any prominent line emission. The aperture extents for the AGN and the optical peak were chosen based on the light profile of the H$\alpha$ and [O{\sc{iii}}]$\lambda$5007\AA \ emission lines in spatial direction, to ensure that we included the total emission of these lines. 

Before we can measure the fluxes of the emission lines, we first need to subtract stellar emission from the extracted spectra. This was done by fitting the emission line-free regions of each spectrum with a combination of a small subset of galaxy template spectra from the \citet{BC03} stellar library. This library consists of model templates for 39 stellar populations with ages between 5.0 $\times$ 10$^{6}$ and 1.2 $\times$ 10$^{10}$ yr, and metallicities between 0.008 and 0.05, with a spectral sampling of 3\AA. This library has been used to fit the continua and measure the emission line fluxes of the Sloan Digital Sky Survey (SDSS) galaxy spectra \citep{Tremonti04}. We performed our spectral synthesis for the three spectra in {\sc{xspec}} v12.8.2,$^{4}$\let\thefootnote\relax\footnote{$^{4}$ The {\sc{XSPEC}} manual is available at http://heasarc.gsfc.nasa.gov/xanadu/xspec/XspecManual.pdf} assuming $z =$ 0.00390, solar metallicity ($Z_{\odot} =$ 0.02), and the \citet{Cardelli89} extinction law. The best-fitted stellar spectra were then subtracted from the observed spectra, which resulted in residual spectra with emission lines only. We then analyzed the detected emission lines using the {\sc{splot}} tasks in {\sc{iraf}}.

\subsection{{\textsl{Gemini}}/T-ReCS}

NGC 1448 was observed at MIR wavelengths in 2010 with high spatial resolution using the Thermal-Region Camera Spectrograph (T-ReCS; field of view 28$\farcs$8 $\times$ 21$\farcs$6; 0.09 arcsec pixel$^{-1}$; \citealp{Telesco98}), mounted on the Gemini-South telescope. The observation was carried out on 2010-08-19 (Program ID GS-2010B-Q-3) for $\approx$319 s on-source time using the $N$-band filter ($\lambda =$ 7.4--13.4 $\mu$m) in parallel chop and nod mode. The data were reduced using the {\sc{midir}} pipeline in {\sc{iraf}} provided by the Gemini Observatory, and the image analysis was performed using the {\sc{idl}} package {\sc{mirphot}}, following \citet{Asmus14}. The data were flux-calibrated with a standard star, HD22663, which was observed immediately before NGC 1448. The resolution of the observation is $\sim$0$\farcs$4 as measured from the FWHM of the standard star.

A compact nucleus is clearly detected in the MIR continuum at position RA $=$ 3:44:31.75 and Dec. $=$ $-$44:38:41.8 (telescope astrometric uncertainty is on the order of 1$\arcsec$) with a SNR $\sim$5. The source appears to be embedded in $\sim$0$\farcs$7 extended emission along the host major axis (position angle PA $\sim$ 44$^{\circ}$). However, owing to the small extent and often unstable point spread function (PSF) in ground-based MIR observations, a second epoch of high spatial resolution MIR images is required in order to confirm this extension. The source is best-fitted with a 2-dimensional Gaussian model (see Figure 3), which measured a flux density of $f_{\rm12 \micron} =$ 12.5 $\pm$ 2.3 mJy. 

\section{X-ray Spectral Fitting}

\subsection{Basic Characterization}

In this section, we describe the broadband X-ray spectral analysis of the AGN in NGC 1448, performed using {\sc{xspec}}. Given the non-negligible contribution of the background flux to the weak source flux, particularly in the \textsl{NuSTAR} data at high energies, we binned both the \textsl{Chandra} and \textsl{NuSTAR} spectra to a minimum of 1 count per bin and optimized the fitting parameters using the Poisson statistic, C-statistic \citep{Cash79}.$^{5}$\let\thefootnote\relax\footnote{$^{5}$ We note that grouping the \textsl{Chandra} and \textsl{NuSTAR} data to a minimum of 20 and 40 counts per bin, respectively, and performing the spectral fits using $\chi^{2}$ minimization yields consistent results with the C-statistic approach.} We included a fixed Galactic absorption component, $N_{\rm{H}}^{\rm{Gal}}$ $=$ 9.81 $\times$ 10$^{19}$ cm$^{-2}$ \citep{Kalberla05}, using the {\sc{xspec}} model ``{\sc{phabs}}" in all spectral fits, and assumed solar abundances for all models. The redshift was fixed at $z =$ 0.00390 in all analyses. We quoted all errors at 90$\%$ confidence, unless stated otherwise. We summarize the main results of the analysis in Table 1, and present the X-ray data and best-fitting models in Figure 4.

We began our spectral modeling by examining the \textsl{NuSTAR} and \textsl{Chandra} data separately. At this point we are not yet considering non-AGN contributions to the X-ray spectrum, such as from the two off-nuclear point sources within the extraction region. We modeled the two spectra in the 3--40 keV and 2--8 keV bands, respectively, using a simple power-law model with Galactic absorption. The best fitting photon indices are very flat: $\Gamma$ $=$ $-$0.29 $\pm$ 0.38 for \textsl{NuSTAR} (C-stat/d.o.f $=$ 229/237), and $\Gamma$ $=$ 0.82$_{-0.77}^{+0.88}$ for \textsl{Chandra} (C-stat/d.o.f $=$ 125/126). A slight excess of counts just above 6 keV was observed in both the \textsl{NuSTAR} and \textsl{Chandra} spectra. This is likely an indication of a fluorescent Fe K$\alpha$ line emission, associated with a spectral component produced by AGN emission that is being reflected off a high column of gas. The fluxes of the two spectra within the common energy range of 3--8 keV measured by the model are \emph{f}$_{3-8}$ $=$ 2.00$_{-0.57}^{+0.23}$ $\times$ 10$^{-14}$ erg s$^{-1}$ cm$^{-2}$ for \textsl{NuSTAR}, and \emph{f}$_{3-8}$ $=$ 3.50$_{-1.34}^{+0.56}$ $\times$ 10$^{-14}$ erg s$^{-1}$ cm$^{-2}$ for \textsl{Chandra}. These fluxes are consistent with each other within the measurement uncertainties, indicating that there was no significant variability between the two observations. 

We then fitted the \textsl{NuSTAR} spectrum simultaneously with the \textsl{Chandra} spectrum between 3 and 40 keV using a simple power-law model and a Gaussian component to model the possible Fe K$\alpha$ emission line, with the line energy and width fixed to $E$ $=$ 6.4 keV and $\sigma$ $=$ 10 eV, respectively. This model measured a photon index of $\Gamma$ $=$ 0.02$_{-0.28}^{+0.27}$ and Fe K$\alpha$ line equivalent width of EW $=$ 2.1$_{-0.8}^{+1.0}$ keV (C-stat/d.o.f $=$ 315/331). The flat photon index and large EW measured for the Fe K$\alpha$ line (EW $\gtrsim$ 1 keV), are characteristic signatures for CT absorption.

To test for this, we model both spectra simultaneously from 3--40 keV with the {\sc{pexrav}} model (\citealp{MZ95}), which has historically been used to model reflection-dominated AGN spectra. This model
produces an AGN continuum which is reflected from a slab-geometry torus with an infinite column density, which is highly unlikely to represent the actual AGN torus. However, it can provide a useful initial test to investigate the AGN spectrum for CT obscuration. Because it does not self-consistently model the fluorescence emission lines expected from a CTAGN, we included a Gaussian component in the model to simulate Fe K$\alpha$ narrow line emission, which is the most prominent line produced by CTAGNs. The centroid energy and width of the line were fixed to $E$ $=$ 6.4 keV and $\sigma$ $=$ 10 eV, respectively. Due to the limited number of counts, the inclination of the reflector was fixed to the default value set by the model; i.e., $\theta_{\rm{inc}}$ $\approx$ 63$^{\circ}$. However, we note that fixing $\theta_{\rm{inc}}$ to other values (e.g., near the lower and upper limits of the model; $\theta_{\rm{inc}}$ $\approx$ 26$^{\circ}$ and 84$^{\circ}$, respectively) have insignificant effects on the parameters obtained. We also fixed the reflection scaling factor to \emph{R} $=$ $-$1, to simulate a pure reflection spectrum. In addition, we included the absorbed transmitted component of the AGN, simulating the Compton scattering and photoelectric absorption by the torus using {\sc{cabs}} and {\sc{zphabs}} models, respectively. This model provides a decent fit to the data (C-stat/d.o.f $=$ 318/330), and measured a column density of $N_{\rm{H}}$(los) $=$ 2.7$^{+u}_{-1.2}$ $\times$ 10$^{24}$ cm$^{-2}$, consistent with CT column.$^{6}$\let\thefootnote\relax\footnote{$^{6}$ ``$u$'' is unconstrained.} The intrinsic photon index inferred from the best-fit model is $\Gamma$ $=$ 1.57 $\pm$ 0.35, in agreement with the typical intrinsic value for an AGN (e.g., \citealp{Burlon11}; \citealp{Corral11}). The reflection component of the model dominates the transmitted component at all spectral energies probed, indicating that the spectra is reflection-dominated, consistent with being heavily obscured. 

\subsection{Physical Modeling}

We proceeded to model the X-ray broadband spectrum of the potentially heavily obscured AGN in NGC 1448 using more physically motivated obscuration models to better characterize the broadband spectrum. The two models used are the {\sc{torus}} model by \citealp{BN11} (Model T), and {MYT\sc{orus}} model by \citealp{MY09} (Model M). Details and results of the two models are described in Sections 3.2.2 and 3.2.3, respectively. In addition to these models, we added extra components required to account for non-AGN contributions to the X-ray spectrum, and to provide a good fit to the data as described below, and in Section 3.2.1.

At low energies, typically at $E$ $\lesssim$ 2 keV, where the direct emission from a heavily obscured AGN is completely absorbed, other processes can dominate. These processes include X-ray emission radiated by unresolved off-nuclear X-ray sources, thermal emission from a hot interstellar medium, gas photoionized by the AGN, and scattered emission from the AGN. Our data are not of sufficient quality to accurately distinguish between these different physical processes. Therefore, we simply parametrized the low energy part of the spectrum covered by \textsl{Chandra} using the thermal model ``{\sc{apec}}" \citep{Smith01},$^{7}$\let\thefootnote\relax\footnote{$^{7}$ We note that the {\sc{apec}} component is mainly contributed by the unresolved emission within the 20$\arcsec$-radius extraction region.} and a power-law component to simulate the scattered emission from the AGN.

A constant parameter, \textsl{C} is often included when analyzing data from multiple X-ray observatories simultaneously to account for the cross-calibration uncertainties between the various instruments, and to account for any significant variability of the targets between the observations. As detailed earlier, we do not find any significant differences between the \textsl{Chandra} and \textsl{NuSTAR} spectra, indicating that there has not been significant variability between the two observations. Based on the \textsl{NuSTAR} calibration paper \citep{Madsen15},  the cross-calibration uncertainty of \textsl{Chandra} with respect to \textsl{NuSTAR} is $\approx$1.1. We therefore decided to fix this parameter to the value found by \citet{Madsen15} as a conservative approach. However, we note that allowing this parameter to vary in all models, returns results that are consistent with this value within the statistical uncertainties. 

Our two models are described in {\sc{xspec}} as follows:\\
\\
Model T $=$ {\sc{constant}} $\ast$ {\sc{phabs}} $\ast$ ({\sc{apec}} \\ \\ $+$ {\sc{cons}} $\ast$ {\sc{zpow}} $+$ 2 $\times$ {\sc{tbabs}} $\ast$ {\sc{zpow}} $+$ \\ \\ {\sc{torus}}), \ \ \ \ \ \ \ \ \ \ \ \ \ \ \ \ \ \ \ \ \ \ \ \ \ \ \ \ \ \ \ \ \ \ \ \ \ \ \ \ \ \ \
\ \ \ \ \ \ \ \ \ \ \ \ \ \ \ \ \ (1)
\\
\\
\\
Model M $=$ {\sc{constant}} $\ast$ {\sc{phabs}} $\ast$ ({\sc{apec}} \\ \\ $+$ {\sc{cons}} $\ast$ {\sc{zpow}} $+$ 2 $\times$ {\sc{tbabs}} $\ast$ {\sc{zpow}} $+$\\ \\ {\sc{zpow}} $\ast$ {\sc{mytz}} $+$ {\sc{myts}} $+$ {\sc{mytl}}). \ \ \ \ \ \ \ \ \ \ \ \ \ \ \ \ \ \ \ \ \ \ \ \  (2)
\\

The two additional absorbed power-law components ({\sc{tbabs}}$\ast${\sc{zpow}}) are included in the models to take into account the two off-nuclear sources detected within the extraction region in the \textsl{Chandra} data. We discuss these sources in the following section.

\begin{figure*}
\epsscale{.51}
%\plotone{pexrav.eps}
\plotone{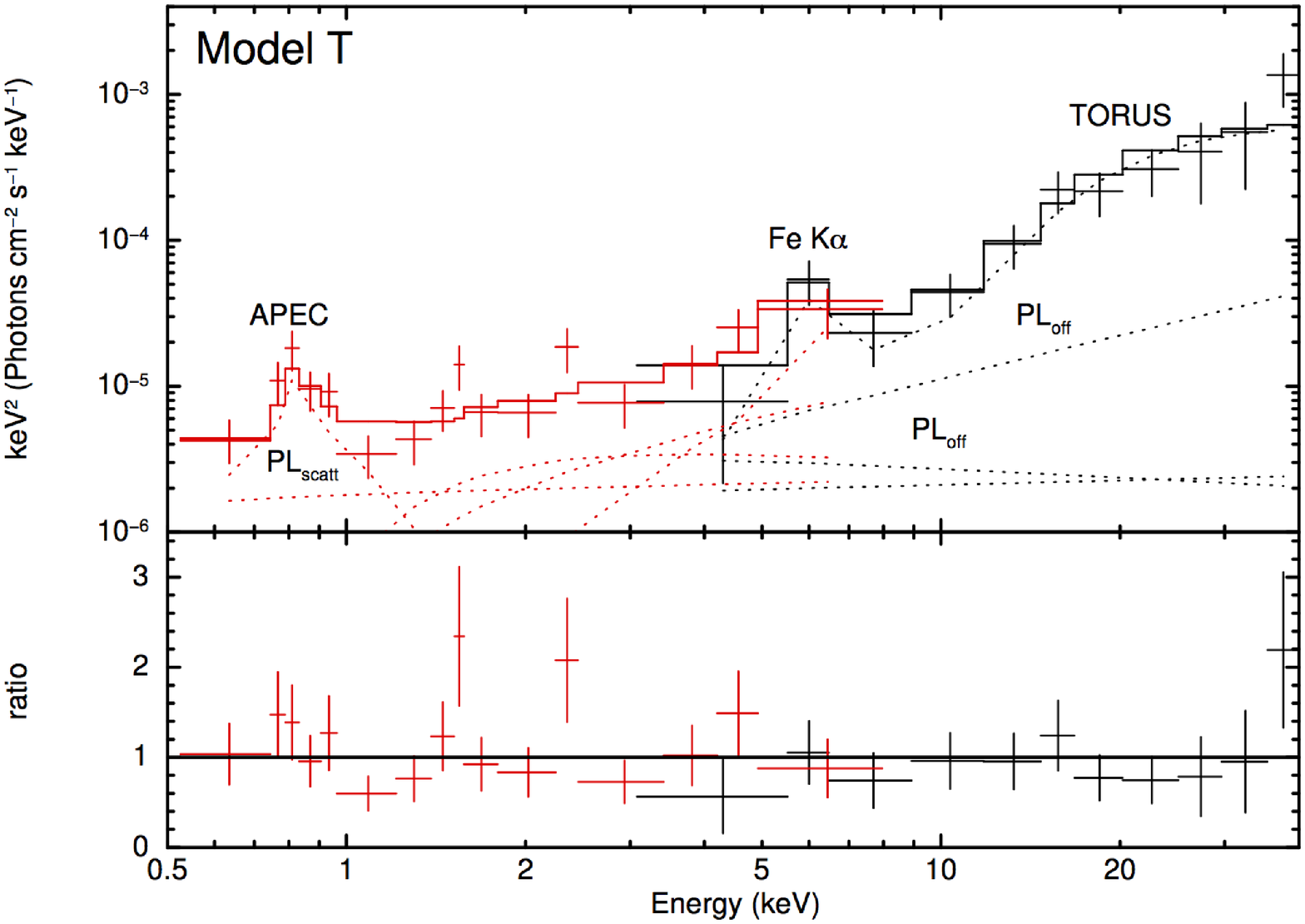}
\epsscale{.51}
\plotone{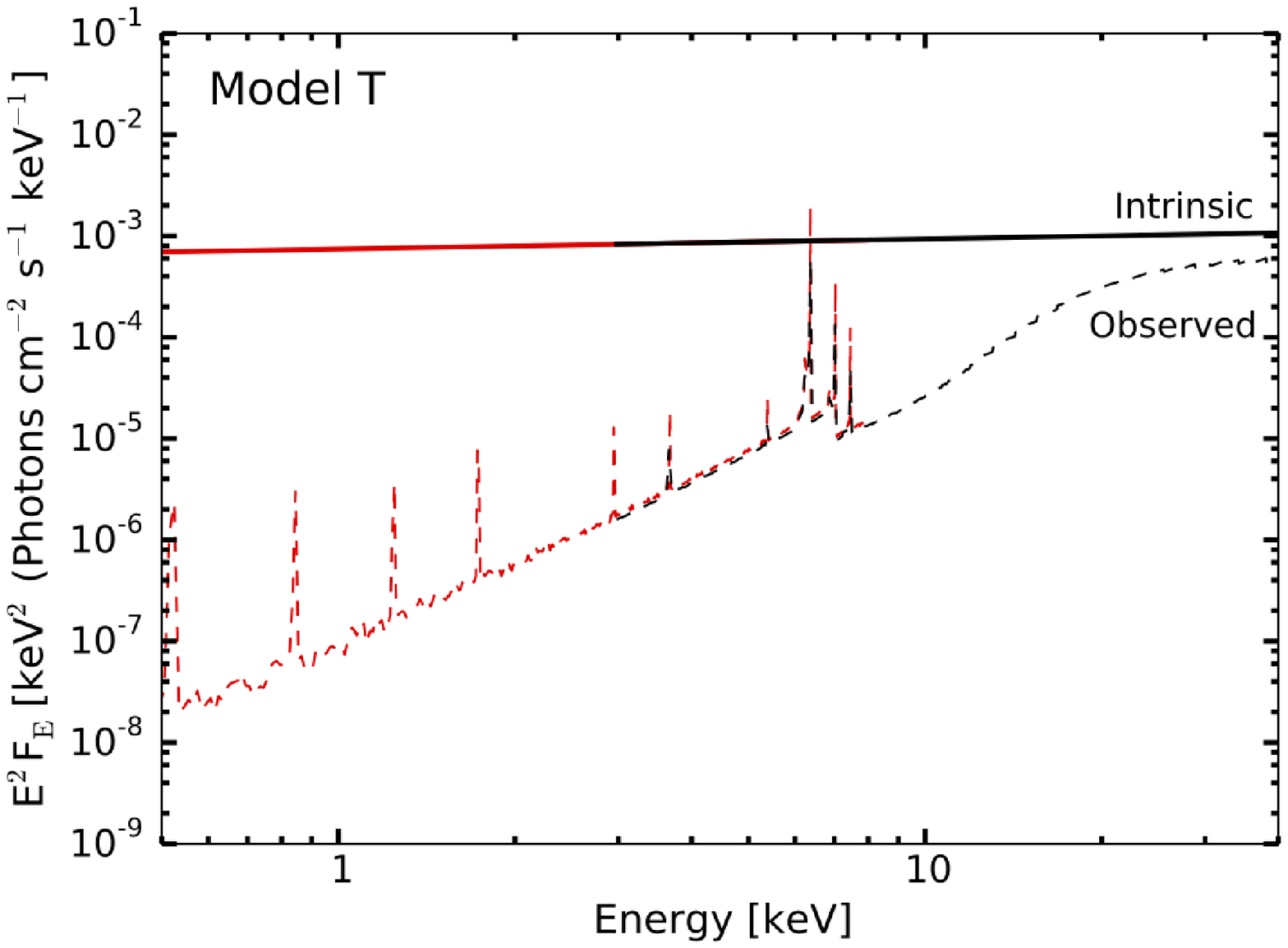}
\epsscale{.51}
\plotone{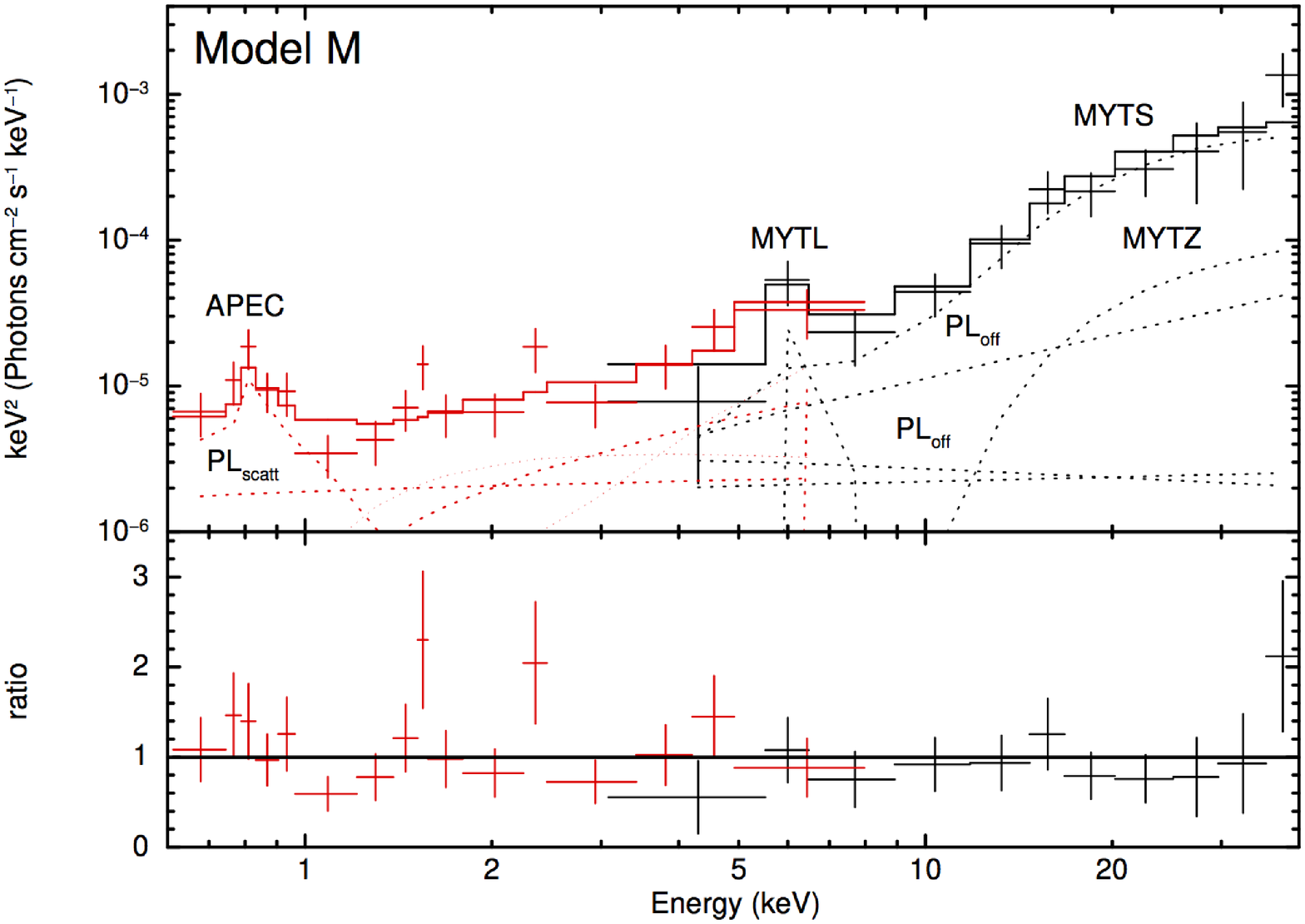}
\epsscale{.51}
\plotone{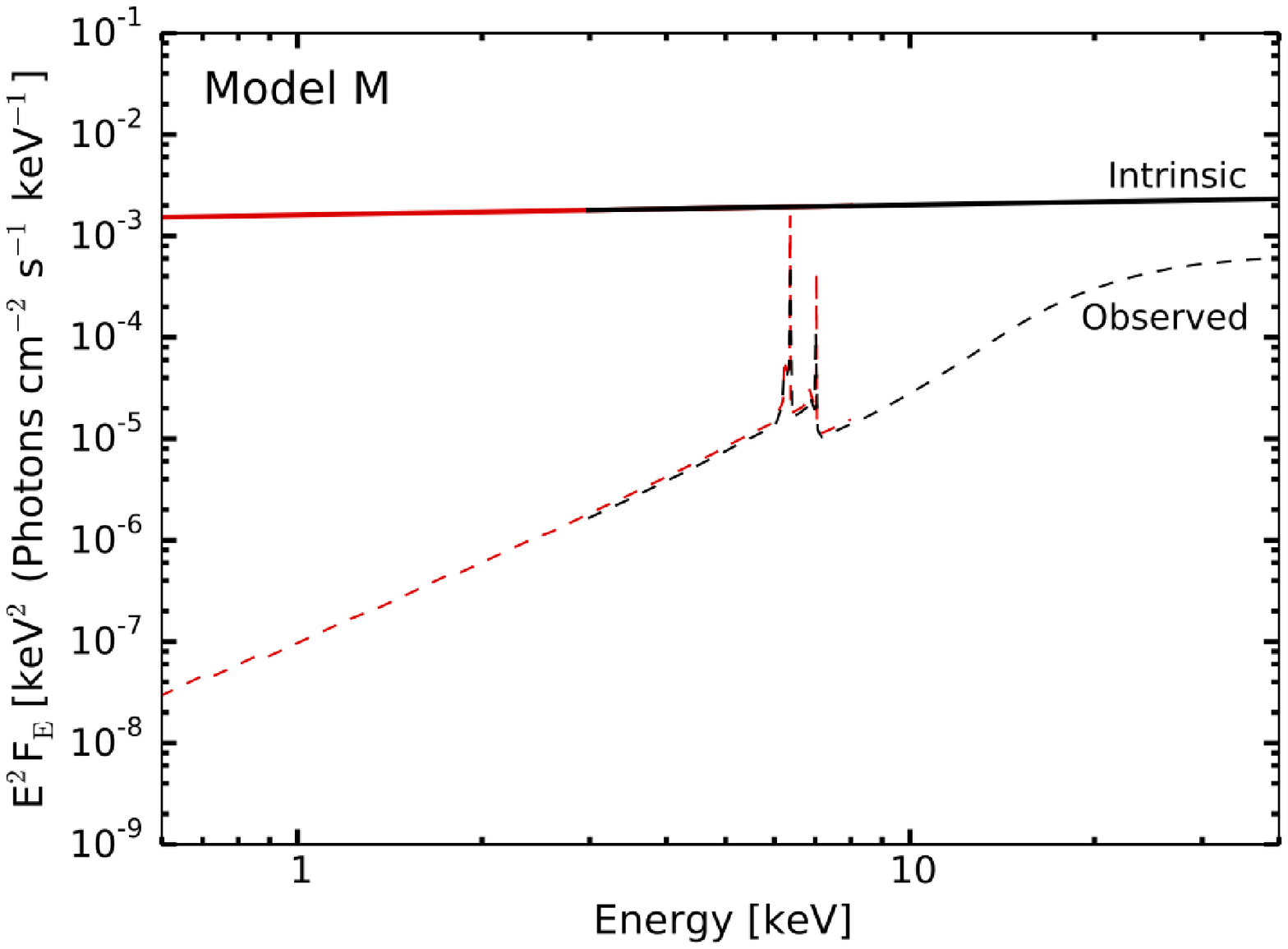}
%\plotone{ironspectrum.eps}
\caption{Best-fitting models to the combined \textsl{NuSTAR} and \textsl{Chandra} data of NGC 1448 - Model T (top) and Model M (bottom). Model T was fitted between 0.5 and 40 keV, and Model M was fitted between 0.6 and 40 keV due to the strong residuals found at $\sim$0.5 keV for this model. The data have been rebinned to a minimum of 3$\sigma$ significance with a maximum of 25 and 100 bins for \textsl{NuSTAR} and \textsl{Chandra}, respectively, for visual clarity. Color scheme: black (\textsl{NuSTAR} FPMA+B), red (\textsl{Chandra}). Plots on the left show the model components fitted to the data (dotted lines), with the total model shown in solid lines. We included an {\sc{apec}} component and a scattered power-law component to model the emission at the softest energies, and two power-law components to model the two off-nuclear sources located within the extraction region in all models. The iron line modeled in Model T is labeled as ``Fe K$\alpha$'', and the direct, scattered and line components of Model M are labeled as {\sc{mytz}}, {\sc{myts}} and {\sc{mytl}}, respectively. The top panels of the left plots show the data and unfolded model in $E^{2}F_{E}$ units, while the bottom panels show the ratio between the data and the folded model. Plots on the right show the observed (dashed lines) and intrinsic spectra (solid line) of the AGN component for each model. The slight offset between the red and black lines are due to the cross-calibration uncertainties between \textsl{Chandra} and \textsl{NuSTAR} \citep{Madsen15}. These plots show that even at $E \geq$ 10 keV, the spectra that we observed for CTAGNs are still significantly suppressed (up to $\sim$2 orders of magnitude for the case of NGC 1448), demonstrating the extreme of CT absorption.}
\end{figure*}

\begin{table*}
\begin{center}
\caption{X-ray spectral fitting results for NGC 1448.}
\begin{tabular}{lcccc}
\hline \hline
Component & Parameter & Units  & Model T & Model M \\
\ \ \ \ \ \ (1) & (2) & (3) & (4) & (5) \\
\tableline
{\sc{apec}} & $kT$ & keV & 0.63$^{+0.14}_{-0.33}$ & 0.63$^{+0.14}_{-0.42}$  \\
Absorber/Reflector & $N_{\rm{H}}$(eq) & 10$^{24}$ cm$^{-2}$ & 4.2$^{+u}_{-1.7}$ & 4.9$^{+u}_{-2.0}$  \\
                   & $N_{\rm{H}}$(los) & 10$^{24}$ cm$^{-2}$ & 4.2$^{+u}_{-1.7}$ & 4.5$^{+u}_{-1.8}$  \\
                   & $\theta_{\rm{inc}}$ &  deg & 87$^{f}$ & 78.6$^{+6.5}_{-11.8}$   \\
                   & $\theta_{\rm{tor}}$ &  deg & 45.9$^{+33.1}_{-18.9}$ & 60$^{f}$  \\
AGN Continuum   & $f_{\rm{scatt}}$   & \% & 0.2$^{+0.3}_{-0.2}$    &  0.1$^{+0.2}_{-0.1}$  \\
                   & $\Gamma$ & &  1.9$^{f}$ & 1.9$^{f}$   \\
                   & \emph{L}$_{0.5-2\rm{,obs}}$ & 10$^{39}$ erg s$^{-1}$   & 0.2 & 0.2  \\
                   & \emph{L}$_{2-10\rm{,obs}}$ & 10$^{39}$ erg s$^{-1}$ & 0.9 & 0.9 \\
                   & \emph{L}$_{10-40\rm{,obs}}$ &  10$^{39}$ erg s$^{-1}$  & 11.7 & 11.8 \\
                   & \emph{L}$_{0.5-2\rm{,int}}$ & 10$^{40}$ erg s$^{-1}$   &  2.6 & 5.6  \\
                   & \emph{L}$_{2-10\rm{,int}}$ & 10$^{40}$ erg s$^{-1}$   & 3.5 & 7.6 \\
                   & \emph{L}$_{10-40\rm{,int}}$ &  10$^{40}$ erg s$^{-1}$   & 3.5 & 7.6 \\
C-stat/d.o.f. & & & 431/445 & 429/440  \\           
\tableline
\end{tabular}
\tablecomments{(1) Model component; (2) parameter associated with each component; (3) units of each parameter; (4) best-fitting parameters for Model T ({\sc{torus}} model by \citealp{BN11});  (5) best-fitting parameters for Model M ({MYT\sc{orus}} model by \citealp{MY09}). ``f'' is fixed parameter and ``u'' is unconstrained parameter. Details of each model are described in Section 3.}
\end{center}
\end{table*}

\subsubsection{Off-nuclear X-ray Sources}

In this section, we present the analysis of the two off-nuclear point sources detected in the 0.5--8 keV \textsl{Chandra} band within the 20$\arcsec$-radius extraction region (see Figure 1). We extracted their spectra using extraction regions of 3$\arcsec$ and 2$\arcsec$ for the NE and NW sources, respectively. Due to limited counts ($\sim$30 counts at 0.5--8 keV for each source), we binned their spectra to a minimum of 5 counts per bin and optimized the fitting parameters using C-statistic. We model both sources using a simple power-law model absorbed by host galaxy absorption ({\sc{tbabs}}), in addition to the Galactic absorption.

The photon indices and intrinsic absorption columns measured by the fits are $\Gamma$ $=$ 1.02$_{-1.28}^{+1.84}$ and \emph{N}$_{\rm H}$(\rm int) $\leq$ 2.19 $\times$10$^{22}$ cm$^{-2}$ (C-stat/d.o.f $=$ 8.1/7) for the NE source, and $\Gamma$ $=$ 2.15$_{-1.30}^{+2.03}$ and \emph{N}$_{\rm H}$(\rm int) $\leq$ 2.11 $\times$10$^{22}$ cm$^{-2}$ (C-stat/d.o.f $=$ 7.4/10) for the NW source. The 0.5--8 keV intrinsic luminosities for both sources, assuming that they are located within NGC 1448, are 2.23$_{-0.85}^{+1.83}$ $\times$ 10$^{38}$ erg s$^{-1}$ and 1.56$_{-0.09}^{+1.77}$ $\times$ 10$^{38}$ erg s$^{-1}$ for the NE and NW sources, respectively. These luminosities are almost an order of magnitude lower than the threshold luminosity for ultraluminous X-ray sources (ULXs; $L_{\rm X} \gtrsim$ 10$^{39}$ erg s$^{-1}$). 

The best-fitted photon index for the NE source is potentially flat (although with large uncertainties), and may suggest that it is a background obscured AGN. We estimate the probability of finding an AGN within a random 20$\arcsec$-radius region with at least the observed flux of the NW source ($f_{0.5-8} \gtrsim$ 4 $\times$ 10$^{-15}$ erg s$^{-1}$ cm$^{-2}$) using the AGN number counts of the 4 Ms \textsl{Chandra} Deep Field South (CDFS) Survey \citep{Lehmer12}. Based on this, the probability of finding two or more background AGNs within a 20$\arcsec$-radius circular region is $<$ 1$\%$. In addition, we do not find counterparts to these \textsl{Chandra} sources at other wavelengths (e.g., MIR and optical). Given their high Galactic latitudes ($\mid$$b$$\mid$ $\sim$ 51$^{\circ}$), it is unlikely that these are Galactic sources. Based on these arguments, we conclude that the two off-nuclear X-ray sources are more likely to be X-ray binaries within NGC 1448. Given the stellar mass and SFR of the galaxy, we would expect to find $\sim$3 X-ray binaries with $L_{0.5-8} >$ 10$^{38}$ erg s$^{-1}$ in NGC 1448 (\citealp{Gilfanov04}; \citealp{Mineo12}). This is consistent with our detections of two potential candidates within the central 20$\arcsec$ circular radius of the galaxy. We therefore included the two absorbed power-law components detailed above into Model T and M to account for their maximum contributions to our X-ray spectra of NGC 1448. We fixed all the parameters at the best-fit values measured from the \textsl{Chandra} data. 

Due to the \textsl{NuSTAR} PSF, sources that lie outside the 20$\arcsec$-radius extraction region could also contaminate our spectrum. We tested whether these sources contribute a significant fraction of our X-ray broadband spectrum by including the best-fit power-law model of the source $\sim$30$\arcsec$ west of the AGN to our X-ray spectral modeling of NGC 1448. This source is the brightest in a 50$\arcsec$-radius circular region around the AGN which corresponds to a $\sim$70$\%$ \textsl{NuSTAR} ECF. We left the normalization of the source component free to vary, and found that the value is consistent with zero, suggesting that the source does not significantly contaminate our broadband AGN spectrum. We therefore did not include this source, and consequently any of the fainter sources outside the extraction region, in our modeling of the AGN X-ray spectra. 

\subsubsection{Model T}

The {\sc{torus}} model by \citet{BN11} (Model T) simulates obscuration by a spherical torus with variable biconical polar opening angle ($\theta_{\rm{tor}}$) ranging between 26--84$^{\circ}$. The line-of-sight column density, $N_{\rm{H}}$(los), through the torus, which is equal to the equatorial column density $N_{\rm{H}}$(eq), is independent of the inclination angle ($\theta_{\rm{inc}}$), and extends up to 10$^{26}$ cm$^{-2}$, an order of magnitude higher than that allowed by the MYT{\sc{orus}} model (see Section 3.2.3). The model also self-consistently predicts the Compton scattering and fluorescent Fe K$\alpha$ and Fe K$\beta$ lines, as well as K$\alpha$ emission from C, O, Ne, Mg, Si, Ar, Ca, Cr, and Ni, which are commonly seen in CTAGNs. The model is available between 0.1 and 320 keV.

In addition to the {\sc{torus}} model, we included several other model components as described earlier in Section 3 and 3.2.1 to provide a good fit to the data. Initially, we left both $\theta_{\rm{inc}}$ and $\theta_{\rm{tor}}$ free to vary; however, they were both unconstrained. We then fixed the torus inclination angle to the upper limit of the model ($\theta_{\rm{inc}}$ $=$ 87$^{\circ}$), to simulate an edge-on inclination torus, and to allow for the full exploration of $\theta_{\rm{tor}}$ \citep{Brightman15}. The column density measurement is insensitive to $\theta_{\rm{inc}}$ when $\theta_{\rm{inc}}$ $>$ $\theta_{\rm{tor}}$ \citep{Brightman15}, and therefore should not significantly affect the $N_{\rm{H}}$(los) measured. In addition, we fixed the photon index to $\Gamma$ $=$ 1.9 as it could not be constrained simultaneously with $\theta_{\rm{tor}}$.

The model implied a column density of $N_{\rm{H}}$(los) $=$ 4.2$^{+u}_{-1.7}$ $\times$ 10$^{24}$ cm$^{-2}$ (C-stat/d.o.f $=$ 431/445). The lower limit of this column density is well above the CT threshold, and therefore this model confirms that the AGN is CT. The best-fit torus opening angle is $\theta_{\rm{tor}}$ = 45.9$^{+33.1}_{-18.9}$ degrees, suggesting a geometrically thick torus. The model measured a small scattering fraction of 0.2$^{+0.3}_{-0.2}\ \%$ with respect to the intrinsic power-law. This is consistent with that found in other obscured AGN (e.g., \citealp{Noguchi10}; \citealp{Gandhi14}; \citealp{Gandhi15}). However we note that given the modest quality of our data, the scattered power-law component will also include contribution from other processes such as unresolved X-ray binaries. Therefore, the true AGN scattering fraction could be lower than this value. We inferred the intrinsic luminosities of the AGN in three different bands based upon the best-fit parameters obtained, as presented in Table 1.

%$\theta_{\rm{tor}}$ to 30, 60 and 80 degrees to simulate a geometrically thick torus, a torus with intermediate opening angle and a geometrically thin torus, respectively, as this parameter could not be constrained due to the modest quality of our data. All three models yielded good fits to the data with reduced $\chi^{2}$ $\sim$ 1. However, only the fit with $\theta_{\rm{tor}}$ $=$ 30$^{\circ}$ gave a constrained photon index. We therefore favor the solution inferred by this geometrically thick torus model. Our best-fit model with $\theta_{\rm{tor}}$ fixed to 30$^{\circ}$ measured a column density of \emph{N}$_{\rm H}$(\rm los) $\gtrsim$ 3.6 $\times$ 10$^{24}$ cm$^{-2}$, with an unconstrained upper limit ($\chi^{2}$/d.o.f $=$ 17.8/17). The photon index inferred is quite flat, $\Gamma$ $=$ 1.52$^{+1.04}_{-0.37}$, although consistent with the typical value for an AGN. The cross-calibration constant of \textsl{Chandra} with respect to \textsl{NuSTAR} is $\sim$1 within the statistical uncertainties, indicating that the spectra are in good agreement with each other. We inferred the intrinsic luminosities from this model based upon the best-fit parameters obtained, as presented in Table 1.

\subsubsection{Model M}

The MYT{\sc{orus}} model by \citet{MY09} (Model M) simulates a toroidal absorber geometry with a fixed opening angle of $\theta_{\rm{tor}}$ $=$ 60$^{\circ}$ and variable inclination angle. The line-of-sight column density, $N_{\rm{H}}$(los), is derived from the measured inclination angle and equatorial column density, $N_{\rm{H}}$(eq), which is simulated only up to a column density of $N_{\rm{H}}$(eq) $=$ 10$^{25}$ cm$^{-2}$. There is more freedom in exploring complex absorbing geometry in the MYT{\sc{orus}} model as it allows the user to disentangle the direct ({\sc{mytz}}), scattered ({\sc{myts}}) and line-emission ({\sc{mytl}}) components from each other. The {\sc{mytl}} component simulates the neutral Fe K$\alpha$ and Fe K$\beta$ fluorescence lines, and the associated Compton shoulders for the AGN. The model is defined between 0.5 and 500 keV. However, we noticed strong residuals at $\sim$0.5 keV which could be attributed to the fact that we are probing the lower limit of the model. Therefore for this model, we restricted our fit to above 0.6 keV.

For simplicity, we fitted the AGN spectrum by coupling all the parameters of the scattered and fluorescent line components to the direct continuum component. The relative normalizations of {\sc{myts}} ($A_{S}$) and {\sc{mytl}} ($A_{L}$) with respect to {\sc{mytz}} ($A_{Z}$) were set to 1. At first, we left the photon index free to vary; however, it reached the lower limit of the model, suggesting that it is not well constrained. Therefore, we fixed the value of this parameter to $\Gamma$ $=$ 1.9.

Model M also gives as good a fit to the data as Model T, with C-Stat/d.o.f $=$ 429/440. Using this model, we measured an equatorial column density of $N_{\rm{H}}$(eq) $=$ 4.9$^{+u}_{-2.0}$ $\times$ 10$^{24}$ cm$^{-2}$. The model measured a high inclination angle of $\theta_{\rm{inc}}$ $=$ 78.6$^{+6.5}_{-11.8}$ degrees, close to the maximum value fixed in Model T. The corresponding \emph{N}$_{\rm H}$(\rm los) value is well within the CT regime, within the uncertainties, \emph{N}$_{\rm H}$(\rm los) $=$ 4.5$^{+u}_{-1.8}$ $\times$10$^{24}$ cm$^{-2}$, and agrees very well with that measured by Model T.$^{8}$\let\thefootnote\relax\footnote{$^{8}$ The line-of-sight column density for Model M was derived using equation 3.1 of \citet{MY09}, and assuming an inclination angle at the best-fit value.} This model also measured a small scattering fraction; i.e., 0.1$^{+0.2}_{-0.1}\ \%$, consistent with Model T. As with Model T, we determined the intrinsic luminosity of the AGN in different bands from the best-fitting parameters (see Table 1).

\section{Multiwavelength Results}

In this section, we present the results from the optical and MIR observations of NGC 1448 as detailed in Section 2.3 and 2.4, respectively. We use these data to calculate the properties of the AGN at different wavelengths to compare with the results of our X-ray spectral analysis.

\subsection{Optical}

As described in Section 2.3, we extracted optical spectra from three different regions along the slit; i.e, the AGN, optical peak and across the whole galaxy. The extracted spectra, along with the fits obtained from the spectral synthesis modeling (see Section 2.3), and the fit residuals, are shown in Figure 5. As shown in this figure, no significant residuals were left after fitting the observed spectra with stellar population templates, without the need of additional AGN continuum components. This suggests that the AGN optical continuum is highly obscured. The spectral analysis revealed that the stellar emission from all three regions is due to both young (5 Myr) and old (5 Gyr) stellar populations, dominated by the latter with $\sim$75--84$\%$ contributions measured between the three spectra. Although the optical peak is where the total optical emission is the brightest, the emission lines are the strongest at the \textsl{Chandra} position, providing independent evidence of the AGN location within the galaxy. 

The interstellar extinction measured towards the stellar population at the AGN position from the spectral synthesis modeling is much higher than at the optical peak position, $A_{V}^{\rm AGN} =$ 2.15$^{+0.16}_{-0.06}$ mag and $A_{V}^{\rm peak} =$ 0.59$^{+0.04}_{-0.05}$ mag, respectively, further suggesting that the optical emission from the AGN is obscured along our line-of-sight. The total extinction measured from the whole galaxy spectra, $A_{V}^{\rm gal} =$ 0.73 $\pm$ 0.06 mag, is $\sim$3$\times$ lower than that measured at the AGN position, demonstrating that the extinction across the galaxy is non-uniform and inhomogeneous.

 \begin{figure*}
    \centering 
    \includegraphics[scale=0.8]{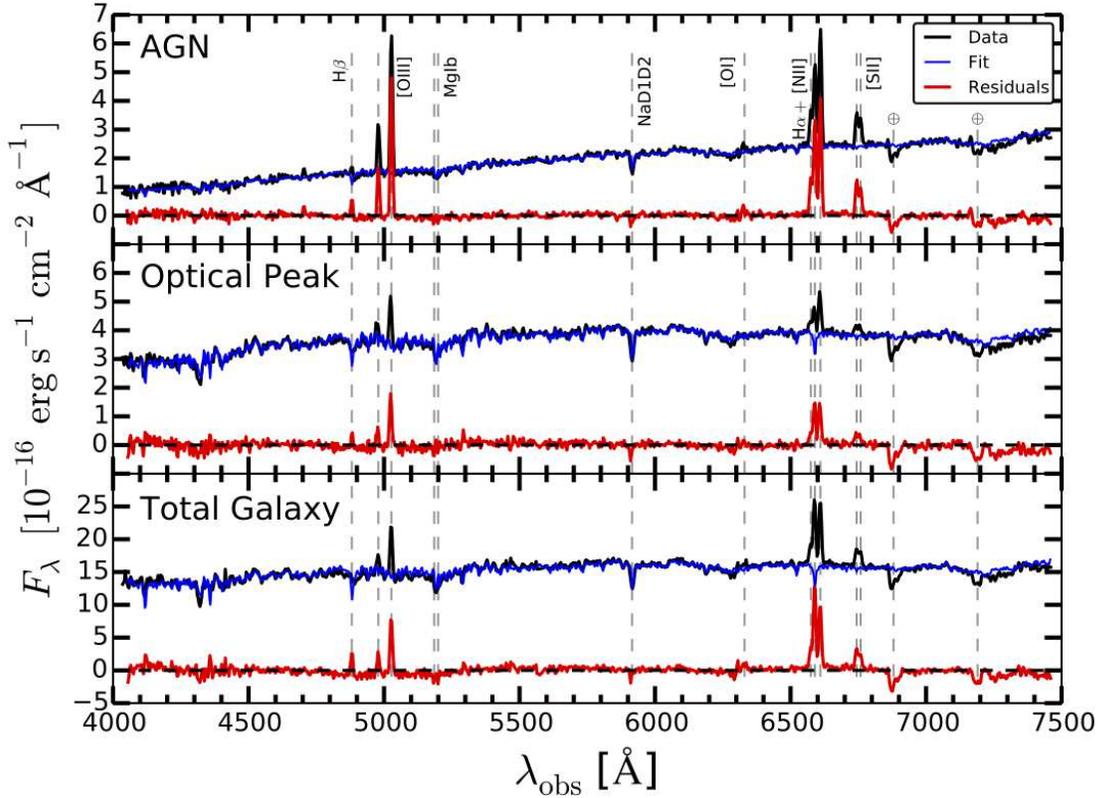}
    \caption{The optical spectra for the AGN (top), optical peak (middle) and the whole galaxy (bottom) extracted from the aperture regions shown in Figure 2. Black are the observed spectra, blue are the best-fitted stellar template spectra, and red are the residual spectra obtained after subtracting the best-fitted stellar template spectra from the observed spectra. Detected emission and absorption lines are labeled with dotted lines. The absorption lines labeled with $\oplus$ are Telluric.}
\end{figure*}

We measured the emission line fluxes and ratios for the AGN, optical peak and the whole galaxy from the extinction-corrected residual spectra (see Table 2). Firstly, we calculated the extinctions toward the AGN narrow line region (NLR) from the Balmer decrements at the three positions. We assumed an intrinsic Balmer decrement of (H$\alpha$/H$\beta$)$_{\rm int}$ $=$ 2.86, which corresponds to a temperature of $T =$ 10$^{4}$ K and an electron density $n_{e} =$ 10$^{2}$ cm$^{-3}$ for Case B recombination \citep{Osterbrock89}. Based on these, we found $A_{V}^{\rm AGN} =$1.89 $\pm$ 0.03 mag, $A_{V}^{\rm peak} =$ 0.74 $\pm$ 0.09 mag, and $A_{V}^{\rm gal} =$ 1.03 $\pm$ 0.04 mag. The high extinction measured at the AGN position provides strong evidence that the AGN is heavily obscured at optical wavelength by the host galaxy.

Using the emission line ratios tabulated in Table 2, we constructed the Baldwin-Philips-Terlevich (BPT) diagnostic diagrams \citep{Baldwin81} for NGC 1448. This is shown in Figure 6. Based on these diagrams, all three spectra fall within the region of Seyfert galaxies, with the AGN having slightly higher emission line ratios. These results provide the first identification of the AGN in NGC 1448 at optical wavelengths, and also confirm the position of the AGN in the galaxy. The total galaxy spectrum is located close to the H{\sc{ii}}/low ionization emission line regions (LINERs) of the BPT diagrams. This might explain why the AGN was misidentified as an H{\sc{ii}} region in \citet{Veron86}; i.e., due to host galaxy contaminations as a result of the larger slit size used to extract the spectrum. 

%{\bf{\textcolor{red}{[Goulding: Could variability of the accretion rate explain the discrepancy between the Veron-Cetty optical spectrum and your new spectrum? It seems odd to me that there is no real explanation at this present time why you would now classify the optical spectrum as an AGN, irrespective of the extraction center, when the Veron-Cetty spectrum looks very different from what we see now.]}}}

The separation between the AGN and the optical peak, $\sim$225 pc, is consistent with the NLR scale observed in AGNs with comparable luminosities as NGC 1448 \citep{Bennert06}. The apparent separation observed between the two can be attributed to the presence of a dust lane which obscures part of the system along our line-of-sight.

The intrinsic [O{\sc{iii}}]$\lambda$5007\AA\ flux of the AGN corrected for the Balmer decrement is about an order of magnitude higher than the observed luminosity (see Table 2). The [O{\sc{iii}}] line emission in AGN is mostly produced in the NLR due to photoionization from the central source. Since this region extends beyond the torus, it does not suffer from nuclear obscuration like the X-ray emission. However as we demonstrated earlier, it can suffer from significant optical extinction from the host galaxy. Indeed, in extreme cases, the host galaxy obscuration can be so high that the optical Balmer decrement only provides a lower limit on the extinction \citep{Goulding09}. 

Using our intrinsic [O{\sc{iii}}] luminosity measured for the AGN, \emph{L}$_{[\rm OIII]}$ $=$ (6.89 $\pm$ 0.08) $\times$ 10$^{38}$ erg s$^{-1}$, we proceeded to compare the intrinsic X-ray luminosity determined from our X-ray spectral fitting with that predicted from the X-ray:[O{\sc{iii}}] intrinsic luminosity relationship of \citet{Panessa06}. Based on this correlation, we infer an intrinsic X-ray luminosity of \emph{L}$_{2-10\rm{,int}}$ $=$ (0.2--4.5) $\times$ 10$^{40}$ erg s$^{-1}$ (see footnote 9). This is consistent with that determined from our X-ray spectral fittings, providing confidence that our analysis is reliable.{\let\thefootnote\relax\footnote{$^{9}$ The given luminosity range accounts for the mean scatter of the correlation.}}

\begin{figure*}
\epsscale{0.45}
\plotone{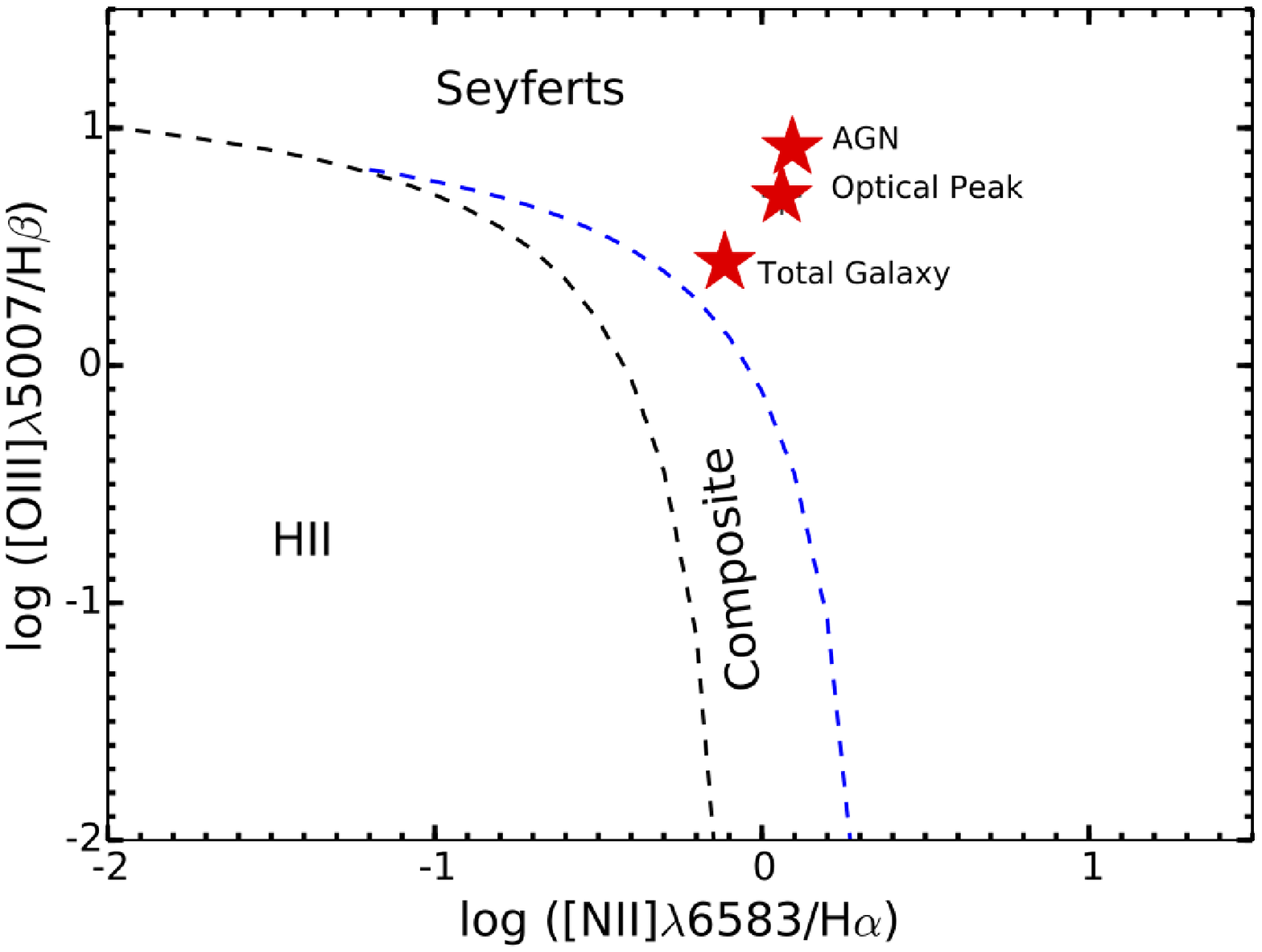}
\epsscale{0.45}
\plotone{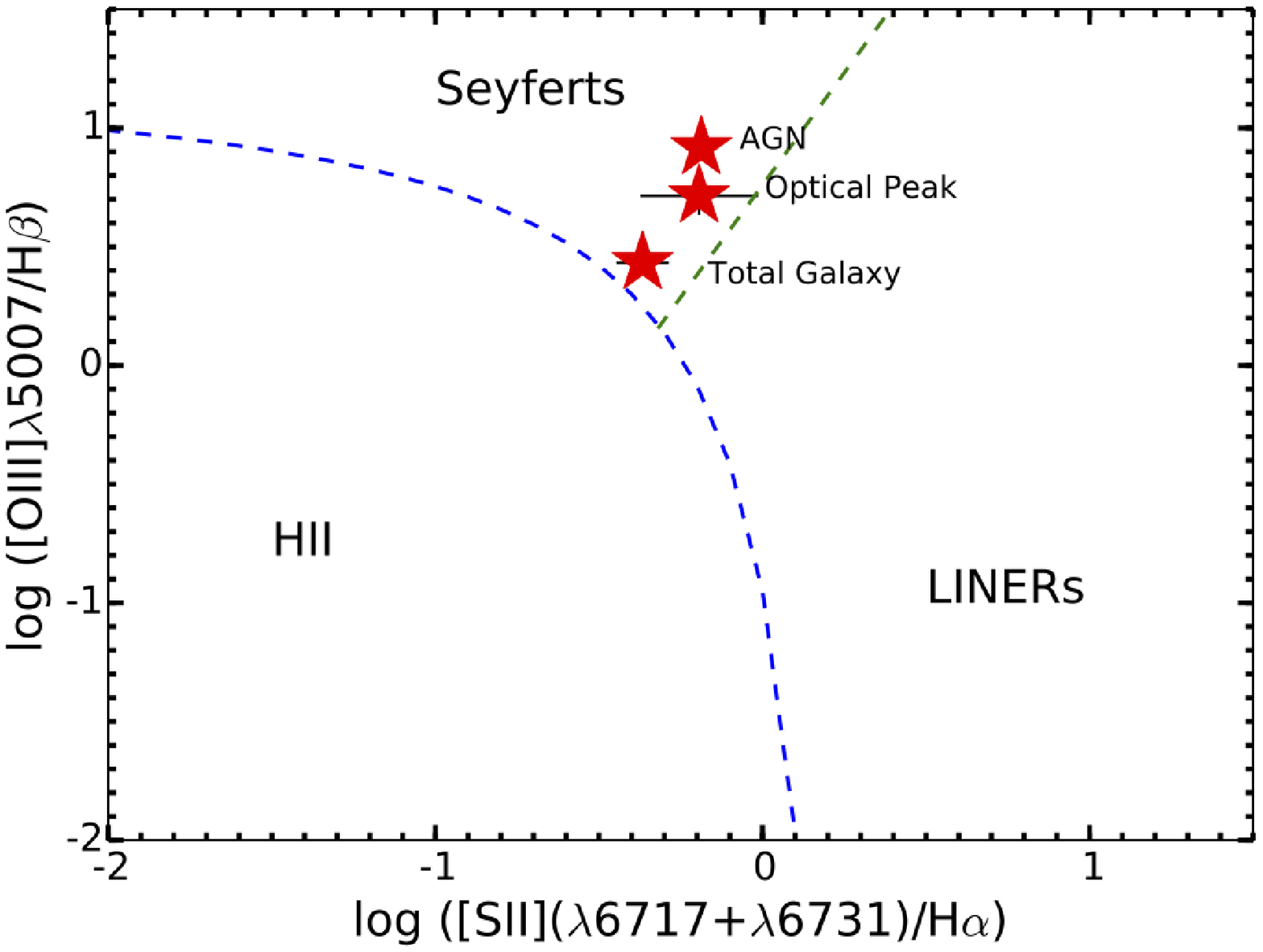}
\epsscale{0.45}
\plotone{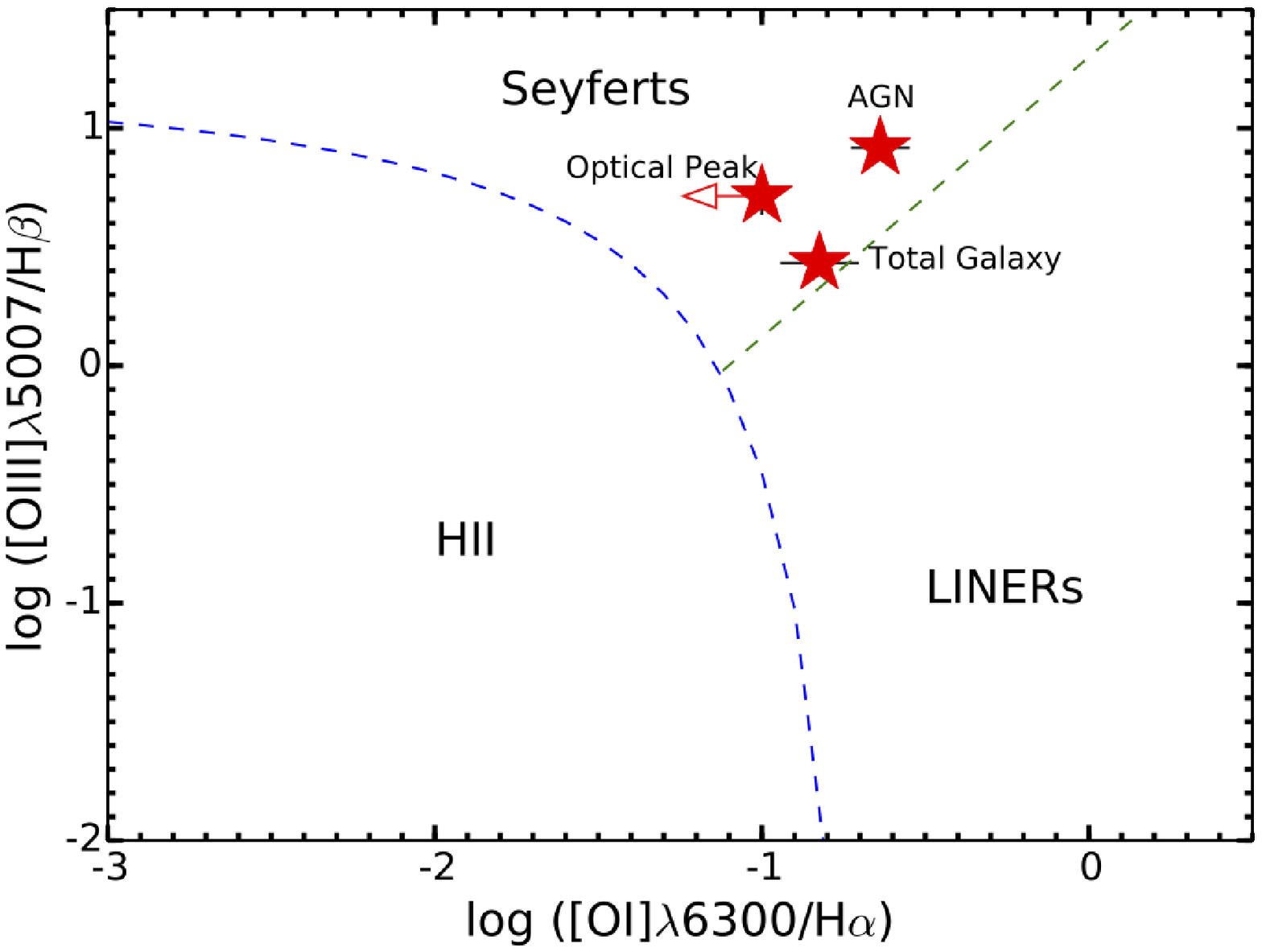}
\caption{The location of the AGN, optical peak and the total galaxy in the BPT diagnostic diagrams. The blue dashed lines indicate the maximum limit for the ionization of gas by a starburst found by \citet{Kewley01}, black dashed lines are the empirical division between H{\sc{ii}} regions and AGNs proposed by \citet{Kauffmann03}, and the green dashed lines are the division between LINERs and Seyfert galaxies, proposed by \citet{Kewley06}. All three regions from our narrow slit spectroscopy show evidence for an AGN in NGC 1448.}
\end{figure*}

\begin{table}
\begin{center}
\caption{Optical emission line fluxes and ratios for spectra shown in Figure 5.}
\begin{tabular}{lccc}
\hline \hline
Emission Line & AGN & Optical  & Total   \\
                &          &   Peak & Galaxy \\
\tableline
H$\alpha$ & 3.79 $\pm$ 0.08 & 1.51 $\pm$ 0.17 & 15.2 $\pm$ 0.50 \\
H$\beta$ & 0.65 $\pm$ 0.05 & 0.40 $\pm$ 0.07 & 3.60 $\pm$ 0.32 \\
$[$O{\sc{iii}}]$\lambda$4959 & 1.85 $\pm$ 0.05 & 0.74 $\pm$ 0.10 & 3.51 $\pm$ 0.38 \\
$[$O{\sc{iii}}]$\lambda$5007 (obs.)& 5.38 $\pm$ 0.06 & 2.07 $\pm$ 0.09 & 9.78 $\pm$ 0.37 \\
$[$O{\sc{iii}}]$\lambda$5007 (int.)& 43.7 $\pm$ 0.49 & 16.8 $\pm$ 0.73 & 79.4 $\pm$ 3.00 \\
$[$O{\sc{i}}]$\lambda$6300 & 0.38 $\pm$ 0.08 & $<$ 0.34  & 2.22 $\pm$ 0.60 \\
$[$N{\sc{ii}}]$\lambda$6549 & 1.51 $\pm$ 0.08 & 0.58 $\pm$ 0.19 & 3.44 $\pm$ 0.55 \\
$[$N{\sc{ii}}]$\lambda$6583 & 4.71 $\pm$ 0.06 & 1.73 $\pm$ 0.10 & 11.7 $\pm$ 0.39 \\
$[$S{\sc{ii}}]$\lambda$6717 & 1.36 $\pm$ 0.09  & 0.40 $\pm$ 0.27 & 3.41 $\pm$ 0.87 \\
$[$S{\sc{ii}}]$\lambda$6731 & 1.12 $\pm$ 0.09 & 0.57 $\pm$ 0.29 & 3.14 $\pm$ 0.93 \\
\hline \hline
Line Ratio & & & \\
\tableline
H$\alpha$/H$\beta$ & 5.83 $\pm$ 0.47 & 3.78 $\pm$ 0.79 & 4.22 $\pm$ 0.40 \\
$[$O{\sc{iii}}]$\lambda$5007/H$\beta$ & 8.28 $\pm$ 0.64 & 5.18 $\pm$ 0.93 & 2.72 $\pm$ 0.26 \\
$[$O{\sc{i}}]$\lambda$6300/H$\alpha$ & 0.10 $\pm$ 0.02 & $<$ 0.23 & 0.15 $\pm$ 0.04 \\
$[$N{\sc{ii}}]$\lambda$6583/H$\alpha$ & 1.24 $\pm$ 0.03 & 1.15 $\pm$ 0.15 & 0.77 $\pm$ 0.04 \\
 $[$S{\sc{ii}}]($\lambda$6717$+\lambda$6731)/H$\alpha$ & 0.65 $\pm$ 0.04 & 0.64 $\pm$ 0.27 & 0.43 $\pm$ 0.08 \\
\tableline
\end{tabular}
\tablecomments{The fluxes are given in units of 10$^{-15}$ erg s$^{-1}$ cm$^{-2}$. The intrinsic [O{\sc{iii}}] $\lambda$5007 flux was corrected for the Balmer decrement.}
\end{center}
\end{table}

\begin{figure}
\epsscale{1.2}
\plotone{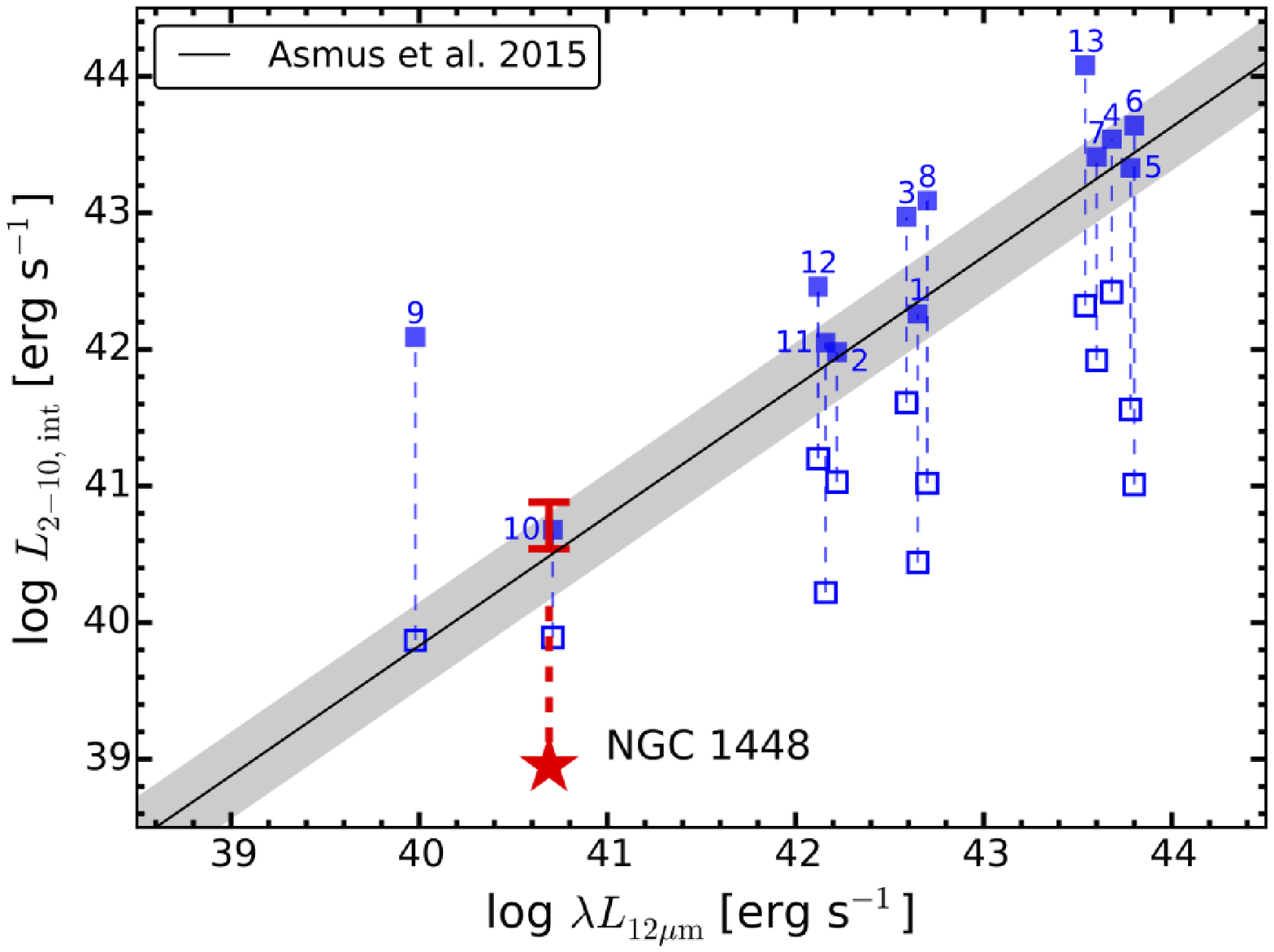}
\caption{The 2--10 keV vs. 12$\mu$m luminosity plot of NGC 1448. The observed 2--10 keV luminosity is plotted using the red star, and the red solid line marks the range of the intrinsic luminosity measured from our X-ray spectral analysis. The black solid line is the 2--10 keV vs. 12$\mu$m luminosity relationship of AGN from \citet{Asmus15}. The shaded region indicates the intrinsic scatter of the relation. The plot shows that our measured intrinsic 2--10 keV luminosity agrees very well with the \citet{Asmus15} correlation, providing evidence that our spectral analysis reliably characterizes the intrinsic power of the AGN. The observed and intrinsic luminosity of the \citet{Gandhi14} local bona-fide CTAGNs with high spatial resolution 12$\mu$m available from \citet{Asmus14}, are shown in open and filled blue squares, respectively.$^{9}$ As can be seen from this plot, NGC 1448 is consistent with the other bona-fide CTAGNs with a typical observed 2--10 keV luminosities that are $\sim$2 orders of magnitude lower than the \citet{Asmus15} relation. However, when corrected for absorption, their luminosities are more in agreement with the correlation, with the exception for NGC 4945, which is known to be overluminous in X-rays. The local CTAGNs plotted in the figure are as follows, with the reference for X-ray data given in parentheses if they were not obtained from \citet{Asmus15}: 1) Circinus; 2) ESO5-G4 \citep{Ueda07}; 3) ESO138-G1; 4) Mrk 3; 5) NGC424; 6) NGC 1068; 7) NGC 3281; 8) NGC 3393; 9) NGC 4945; 10) NGC 5194 (\citealp{Terashima98}; \citealp{Goulding12}); 11) NGC 5643 \citep{Annuar15}; 12) NGC 5728; 13) NGC 6240.}
\end{figure}

\subsection{Mid-Infrared}

The high spatial resolution MIR observation by Gemini/T-ReCS measured a 12$\mu$m luminosity of $\lambda L$$_{\rm12\micron}$ $=$ (4.90 $\pm$ 0.93) $\times$ 10$^{40}$ erg s$^{-1}$ for NGC 1448 (see Section 2.4). MIR luminosity is predicted to provide an accurate estimate for the intrinsic luminosity of the AGN. This is because, the absorbed X-ray radiation from the central engine are mostly re-emitted in the MIR by the torus. To further test the absorption-corrected X-ray luminosity measured from our spectral analysis, we also compared the luminosity measured from our analysis with that predicted from the X-ray:MIR correlation constructed based upon high angular resolution MIR observations of local Seyferts, which has been shown to trace the intrinsic X-ray luminosity of AGN very well (e.g., \citealp{Horst08}; \citealp{Gandhi09}; \citealp{Asmus15}). The correlation with respect to the recent \citet{Asmus15} relation is shown in Figure 7. The figure shows that the observed 2--10 keV luminosity of the AGN in NGC 1448 is about two orders of magnitude lower than this relation, similar to other local bona-fide CTAGNs \citep{Gandhi14}, indicating extreme obscuration.$^{10}$\let\thefootnote\relax\footnote{$^{10}$ We note that the bona-fide CTAGN sample \citep{Gandhi14} plotted in Figure 7, 8 and 9 was updated to include a new local bona-fide CTAGN, NGC 5643 \citep{Annuar15}, and exclude NGC 7582 which has been shown to have variable \emph{N}$_{\rm H}$ \citep{Rivers15}.} However, after correcting for absorption measured from our X-ray spectral modeling, the luminosity agrees very well with the intrinsic relationship of \citet{Asmus15}, further supporting our analysis.

\section{Discussion}

\begin{figure}
\epsscale{1.2}
\plotone{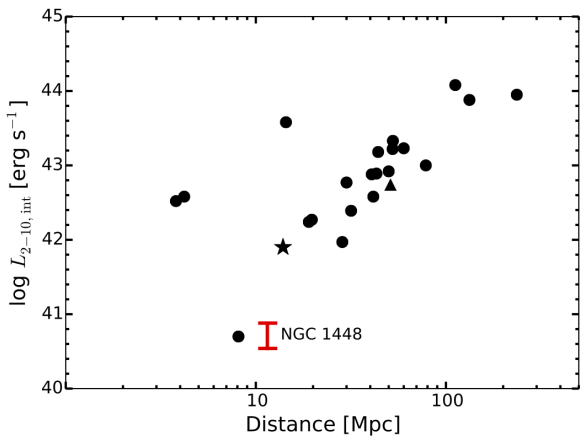}
\caption{The plot of intrinsic 2--10 keV luminosity vs. distance of local bona-fide CTAGN (black circles) updated from \citet{Gandhi14} to include NGC 5643 (black star; \citealp{Annuar15}) and NGC 1448.$^{9}$ Black triangle indicates NGC 4939 which has the lower limit to its 2--10 keV luminosity plotted. The range of intrinsic luminosity of NGC 1448 is plotted in red.}
\end{figure}

In this paper, we presented a multiwavelength view of the AGN in NGC 1448 across three wavebands; i.e., the MIR continuum, optical and X-rays. Our new data provide the first identifications of the AGN at these wavelengths. Combining our \textsl{NuSTAR} data with an archival \textsl{Chandra} observation, we performed broadband X-ray spectral analysis ($\approx$0.5--40 keV) of the AGN. 

We fitted the spectra with two different obscuration models, the {\sc{torus}} and MYT{\sc{orus}} models (Section 3.2.2 and 3.2.3, respectively), which fitted the spectra equally well and gave results that are consistent with each other. Considering both models, we found that the AGN is buried under a CT column of obscuring gas, with a column density of \emph{N}$_{\rm H}$(\rm los) $\gtrsim$ 2.5 $\times$ 10$^{24}$ cm$^{-2}$. Despite the uncertainties in the parameters and the different geometries simulated by the two models, as well as the phenomenological model simulated by the {\sc{pexrav}} model discussed in Section 3.1, a Compton-thick solution is required to get a best-fit to the data in all cases. The 2--10 keV intrinsic luminosity measured from the two physical models range between \emph{L}$_{2-10\rm{,int}}$ $=$ (3.5--7.6) $\times$ 10$^{40}$ erg s$^{-1}$, making NGC 1448 one of the lowest luminosity bona-fide CTAGNs in the local universe, comparable to NGC 5194 (\citealp{Terashima98}; see Figure 8). The measured column density is so extreme that even at $E >$ 10 keV, the observed spectrum is still significantly suppressed (up to $\sim$two orders of magnitude lower than the intrinsic spectrum; see Figure 4). This demonstrates why it is very challenging to identify CTAGNs, even at hard X-ray energies, and at this distance. The predicted 14--195 keV flux of the AGN, \emph{f}$_{14-195} \sim$ 10$^{-12}$ erg s$^{-1}$ cm$^{-2}$, is an order of magnitude lower than the deepest flux level reached by the \textsl{Swift}-BAT all sky survey (\emph{f}$_{14-195} \sim$ 10$^{-11}$ erg s$^{-1}$ cm$^{-2}$; \citealp{Baumgartner13}), demonstrating that pointed observations with \textsl{NuSTAR} have better sensitivity in finding faint CTAGN.

We used our measured 2--10 keV intrinsic X-ray luminosities to estimate the Eddington ratio of NGC 1448. \citet{Goulding10} made an estimation of the supermassive black hole mass ($M_{\rm BH}$) using the $M_{\rm BH}$ to bulge $K$-band luminosity relation by \citet{Marconi03}, and inferred log ($M_{\rm BH}$/$M_{\odot}$) $=$ 6.0$^{+0.1}_{-0.5}$. Based on this black hole mass, we found that the AGN is accreting matter at $\sim$0.6--1.2$\%$ of the Eddington rate.$^{11,12}$\let\thefootnote\relax\footnote{$^{11}$ We assumed an AGN bolometric correction of $L_{\rm bol}/L_{\rm 2-10} =$ 20 (e.g., \citealp{Vasudevan10})}\let\thefootnote\relax\footnote{$^{12}$ The calculation for the Eddington ratio involves highly uncertain quantities, therefore the inferred value is subjected to errors of a factor of a few \citep{Brandt15}.} 

Some studies have suggested that the AGN torus is not developed at low luminosity, $L_{\rm{bol}}$ $\lesssim$ 10$^{42}$ erg s$^{-1}$, and/or at low mass accretion rate, $\lambda_{\rm{Edd}}$ $\lesssim$ 2 $\times$ 10$^{-4}$  (e.g., \citealp{Elitzur06}; \citealp{Hoenig07}; \citealp{Kawamuro16}). The identification of NGC 1448 as a low luminosity CTAGN (as demonstrated from the X-ray spectral fitting and mid-IR analysis) with a bolometric luminosity potentially below the quoted limit, $L_{\rm{bol}} \sim$ (0.7--1.5) $\times$ 10$^{42}$ erg s$^{-1}$ however, does not provide clear evidence for the 
disappearance of the torus at the low AGN luminosities proposed by some studies.

The CTAGN in NGC 1448 is hosted by a nearly edge-on spiral galaxy, which could contribute to at least some of the X-ray obscuration we observed. Our X-ray data are not of sufficient quality to allow us to investigate the origin of the CT obscuration in NGC 1448 in detail. However, our optical spectral synthesis modeling suggests an extinction of $A_{V} =$ 2.15 towards the AGN along our line-of-sight, which translates into a column density of \emph{N}$_{\rm H}$(\rm los) $=$ 4.8 $\times$ 10$^{21}$ cm$^{-2}$, assuming the gas to dust ratio of the Milky Way \citep{Guver09}. This suggests that the CT column density we measured from the X-ray spectral fittings is not due to the host galaxy material, but is due to the AGN circumnuclear material itself. 

%About half of the local bona fide CTAGNs are also located within inclined galaxies ($\theta_{\rm{incl}} \geq$ 60$^{\circ}$; see Figure 10). If the origin of the CT obscurer of these AGNs are in fact the host galaxies, instead of the AGN torus, this could affect our understanding of CTAGNs, which is usually associated with X-ray obscuration by the latter.

\begin{figure}
\epsscale{1.2}
\plotone{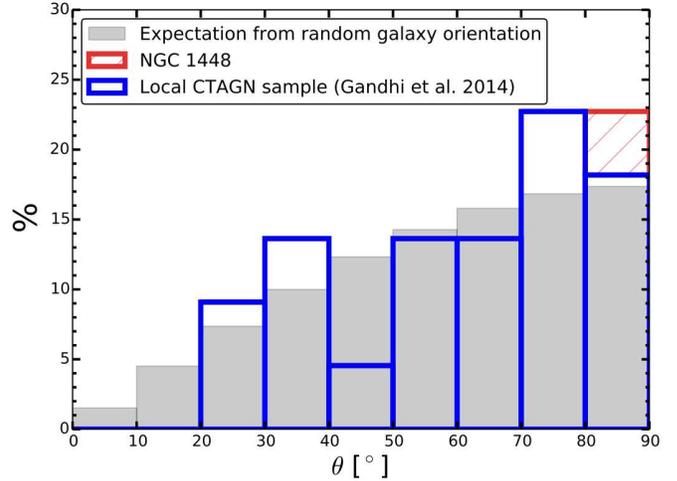}
\caption{Distribution of host galaxy inclination angles ($\theta$) for the \citet{Gandhi14} local CTAGN sample (where available) plotted in blue.$^{9}$ NGC 1448 is plotted in hatched red. The shaded region shows the expected distribution from random galaxy orientation. The total distribution of the local CTAGNs is consistent with the expected random distribution, suggesting that the current sample of local CTAGNs are independent of host galaxy inclinations.}
\end{figure}

High host galaxy inclinations along our line-of-sight can also affect AGN selection in the local universe as demonstrated by, e.g., \citet{Goulding09}. Material in the host galaxies of highly inclined systems can severely suppress optical emission from AGNs, causing them to be misidentified at optical wavelengths. Figure 9 shows the distribution of host galaxy inclinations for local CTAGNs (adapted from \citealp{Gandhi14}),$^{9}$ as compared to the expected distribution from random galaxy orientations. If we are missing CTAGNs in the local universe due to AGN host inclination selection effects, we might expect the local CTAGN sample to be lacking at high host inclinations. However, Figure 9 suggests that this is not the case. Performing a Kolmogorov--Smirnov (KS) test between the two distributions inferred a KS test probability of $P_{\rm KS} \sim$ 0.4. This indicates that the two distributions are not significantly different from each other, and suggests that the current sample of local CTAGNs is not affected by host galaxy inclinations (see also \citealp{Koss16}). Therefore, although the census of CTAGN is incomplete, even in the local universe (at $D$ $\approx$ 200 Mpc), we do not find strong evidence that the incompleteness is due to host galaxy inclination. 

%{\bf{The identifications of NGC 1448 as an AGN in our cosmic backyard for the first time at three different wavebands, calls into question the completeness of AGN samples even in our local universe. \citet{Goulding09} found that $\sim$60$\%$ of their MIR-selected AGN sample are not identified as AGNs in optical (including NGC 1448 from previous optical data; \citealp{Veron86}). In addition, $\sim$50$\%$ of their sample are not detected by the \textsl{Swift}-BAT all sky survey, and therefore not identified as AGNs in X-rays. These facts suggest that we may be missing about half of the AGN population in the local universe, and demonstrate the importance of multiwavelength techniques in AGN identification.}}

\section{Summary}

In this paper, we have presented a multiwavelength view of the AGN in NGC 1448 across MIR continuum, optical, and X-rays. Our data provide the first identifications of the AGN at these wavelengths. The main results can be summarized as follows:

\begin{enumerate}
 \item A broadband ($\approx$0.5--40 keV) X-ray spectral analysis of the AGN using data from \textsl{NuSTAR} and \textsl{Chandra} observations reveals that our direct view towards the AGN is hindered by a CT column of obscuring material, with a column density of \emph{N}$_{\rm H}$(los) $\gtrsim$ 2.5 $\times$ 10$^{24}$ cm$^{-2}$. The range of 2--10 keV intrinsic luminosity measured between the best-fitting torus models is \emph{L}$_{2-10\rm{,int}}$ $=$ (3.5--7.6) $\times$ 10$^{40}$ erg s$^{-1}$, consistent with that predicted from our optical and MIR data. These results indicate that NGC 1448 is one of the lowest luminosity CTAGNs known. The identification of NGC 1448 as a faint CTAGN also demonstrates the depth to which we can find CTAGN with \textsl{NuSTAR} as compared to \textsl{Swift}-BAT. 
 
 \item Optical spectroscopy performed using ESO NTT re-classifed the nuclear spectrum as a Seyfert galaxy based on BPT emission line diagnostics, thus identifying the AGN at optical wavelenghts for the first time.
 
 \item MIR imaging conducted by Gemini/T-ReCS detected a compact nucleus, providing the first identification of the AGN at MIR continuum, further confirming the presence of an AGN in the galaxy. 
 
 \item Comparing the host galaxy inclination distribution of local CTAGNs with that expected from random galaxy orientations, we do not find any evidence that the incompleteness of the current CTAGN census is due to host galaxy inclination effects. 
 
 \end{enumerate}

\section*{Acknowledgments}
We thank the anonymous referee for the useful comments which have helped to improve the paper. We acknowledge financial support from Majlis Amanah Rakyat (MARA) Malaysia (A.A.), the Science and Technology Facilities Council (STFC) grant ST/L00075X/1 (D.M.A.), ST/J003697/1 (P.G.), and ST/K501979/1 (G.B.L.). F.E.B. acknowledges support from CONICYT-Chile (Basal-CATA PFB-06/2007, FONDECYT Regular 1141218, "EMBIGGEN" Anillo ACT1101), and the Ministry of Economy, Development, and Tourism's Millennium Science Initiative through grant IC120009, awarded to The Millennium Institute of Astrophysics, MAS. P.B. would like to thank the STFC for funding. S.M.L.'s research was supported by an appointment to the NASA Postdoctoral Program at the NASA Goddard Space Flight Center, administered by Universities Space Research Association under contract with NASA. M.K. acknowledges support from the Swiss National Science Foundation and Ambizione fellowship grant PZ00P2{\textunderscore}154799/1. We acknowledge financial support from the CONICYT-Chile grants ``EMBIGGEN" Anillo ACT1101 (C.R.), FONDECYT 1141218 (C.R.), Basal-CATA PFB--06/2007 (C.R.) and from the China-CONICYT fund (C.R.).

\textsl{NuSTAR} is a project led by the California Institute of Technology (Caltech), managed by the Jet Propulsion Laboratory (JPL), and funded by the National Aeronautics and Space Administration (NASA). We thank the \textsl{NuSTAR} Operations, Software and Calibrations teams for support with these observations. This research has made use of the \textsl{NuSTAR} Data Analysis Software ({\sc{nustardas}}) jointly developed by the ASI Science Data Center (ASDC, Italy) and the California Institute of Technology (USA). This research also made use of the data obtained through the High Energy Astrophysics Science Archive Research Center (HEASARC) Online Service, provided by the NASA/Goddard Space Flight Center, and the NASA/IPAC extragalactic Database (NED) operated by JPL, Caltech under contract with NASA. 

{\textsl{Facilities}}: {\textsl{Chandra}, {\textsl{Gemini-South}, {\textsl{NTT}, {\textsl{NuSTAR}}.

\bibliographystyle{apj}

\clearpage

\end{document}